%% file: curvevid.tex
\documentclass[a4paper,11pt]{article}

\usepackage{jcappub}
\usepackage{amsmath}
\usepackage{graphicx}
\usepackage{latexsym}
\usepackage{xspace}
\usepackage[table,usenames,dvipsnames]{xcolor}
\usepackage{color}
\usepackage{hyperref} 
\usepackage{bm}
\usepackage{relsize}
\usepackage{multirow}
\usepackage{array}
\usepackage{tabularx}
\usepackage{bayaspic}
\usepackage{multicol}
\usepackage{multirow}
\usepackage{lipsum}
\usepackage{tikz}
\usepackage{fp}

\makeatletter
\gdef\@fpheader{}
\g@addto@macro\bfseries{\boldmath}
\makeatother

\input{newcommands}
\input{tables}

\subheader{}

\title{Inflation with an extra light scalar field \\ after Planck}

\author{Vincent Vennin,}
\author{Kazuya Koyama}
\author{and David Wands}

\affiliation{Institute of Cosmology \& Gravitation, University of Portsmouth, Dennis Sciama Building, Burnaby Road, Portsmouth, PO1 3FX, United Kingdom}

\emailAdd{vincent.vennin@port.ac.uk}
\emailAdd{kazuya.koyama@port.ac.uk}
\emailAdd{david.wands@port.ac.uk}

\date{today}

\begin{document}

\abstract{Bayesian inference techniques are used to investigate situations where an additional light scalar field is present during inflation and reheating. This includes (but is not limited to) curvaton-type models. We design a numerical pipeline where $\simeq 200$ inflaton setups $\times\, 10$ reheating scenarios $= 2000$ models are implemented and we present the results for a few prototypical potentials. We find that single-field models are remarkably robust under the introduction of light scalar degrees of freedom. Models that are ruled out at the single-field level are not improved in general, because good values of the spectral index and the tensor-to-scalar ratio can only be obtained for very fine-tuned values of the extra field parameters and/or when large non-Gaussianities are produced. The only exception is quartic large-field inflation, so that the best models after Planck are of two kinds: plateau potentials, regardless of whether an extra field is added or not, and quartic large-field inflation with an extra light scalar field, in some specific reheating scenarios. Using Bayesian complexity, we also find that more parameters are constrained for the models we study than for their single-field versions. This is because the added parameters not only contribute to the reheating kinematics but also to the cosmological perturbations themselves, to which the added field contributes. The interplay between these two effects lead to a suppression of degeneracies that is responsible for having more constrained parameters.}


\arxivnumber{1512.03403}

\maketitle
\pagebreak

\section{Introduction}
\label{sec:intro}

The recent high-quality measurements~\cite{Adam:2015rua,Ade:2015lrj,Ade:2015ava} of the Cosmic Microwave Background (CMB) anisotropies have shed new light on the physical processes that took place in the very early Universe. These data clearly support  inflation~\cite{Starobinsky:1980te,Sato:1980yn,Guth:1980zm,Linde:1981mu,Albrecht:1982wi,Linde:1983gd} as the leading paradigm for explaining this primordial epoch. At present, the full set of observations can be accounted for in a minimal setup, where inflation is driven by a single scalar inflaton field $\phi$ with canonical kinetic term, minimally coupled to gravity, and evolving in a flat potential $V(\phi)$ in the slow-roll regime. Because inflation proceeds at very high energy where particle physics remains evasive, a variety of such potentials have been proposed in the literature so far. They have been recently mapped in \Ref{Martin:2014vha} where $\sim 80$ potentials are identified and analyzed under the slow-roll framework. The Bayesian evidence of the corresponding $\sim 200$ inflationary models was then derived in \Refs{Martin:2013nzq,Martin:2014rqa,Martin:2014lra,Martin:2014nya}, which was used to identify the best single-field scenarios, mostly of the ``Plateau'' type.

From a theoretical point of view however, as already said, inflation takes place in a regime that is far beyond the reach of accelerators. The physical details of how the inflaton is connected with the standard model of particle physics and its extensions are still unclear. In particular, most physical setups that have been proposed to embed inflation contain extra scalar fields that can play a role either during inflation or afterwards. This is notably the case in string theory models where extra light scalar degrees of freedom are usually considered~\cite{Turok:1987pg, Damour:1995pd, Kachru:2003sx, Krause:2007jk, Baumann:2014nda}. A natural question~\cite{Langlois:2008mn, Clesse:2008pf, Sugiyama:2011jt, Peterson:2011yt, Biagetti:2012xy, Battefeld:2012qx, Levasseur:2013tja, Burgess:2013sla, Turzynski:2014tza, Price:2014xpa} is therefore whether single-field model predictions are robust under the introduction of these additional fields, and whether these fields change the potentials for which the data show the strongest preference.

In this paper, we address this issue using Bayesian inference techniques. We present the results of a systematic analysis of single-field slow-roll models of inflation when an extra light (relative to Hubble scale) scalar field $\sigma$ is introduced and plays a role both during inflation and afterwards. In the limit where this added field $\sigma$ is entirely responsible for the observed primordial curvature perturbations, the class of models this describes is essentially the curvaton scenarios~\cite{Linde:1996gt,Enqvist:2001zp,Lyth:2001nq,Moroi:2001ct,Bartolo:2002vf}. Here however, we address the generic setup where both $\phi$ and $\sigma$ can a priori contribute to curvature perturbations~\cite{Dimopoulos:2003az,Langlois:2004nn,Lazarides:2004we,Moroi:2005np}.\footnote{
In this work, we assume that all particles are in full thermal equilibrium after $\phi$ and $\sigma$ decay; thus there are no residual isocurvature modes~\cite{Lyth:2002my,Weinberg:2004kf}. Any isocurvature modes surviving after reheating would provide very strong additional constraints, but are dependent on the specific process of decay and thermalisation~\cite{Langlois:2004nn, Lemoine:2006sc, Langlois:2008vk, Lemoine:2008qj,Smith:2015bln} which we do not consider here.} In particular, while we require that $\phi$ becomes massive at the end of inflation, we do not make any assumption as to the ordering of the three events: $\sigma$ becomes massive, $\phi$ decays and $\sigma$ decays. Nor do we restrict the epochs during which $\sigma$ can dominate the energy content of the Universe. This leaves us with 10 possible cases (including situations where $\sigma$ drives a secondary phase of inflation~\cite{Langlois:2004nn, Moroi:2005kz, Ichikawa:2008iq, Dimopoulos:2011gb}). These ten ``reheating scenarios'' are listed and detailed in \Ref{Vennin:2015vfa} but, for convenience, they are sketched in the \Fig{fig:rho} of appendix~\ref{sec:ReheatingCases}. The usual curvaton scenario corresponds to case number 8 but one can see that a much wider class of models is covered by the present analysis.

An important aspect of this work is also that reheating kinematic effects are consistently taken into account. In practice, this means that the number of \efolds elapsed between the Hubble exit time of the CMB pivot scale and the end of inflation is not a free parameter but is given by an explicit function of the inflaton potential parameters, the mass of the extra scalar field, its \textit{vev} at the end of inflation and the decay rates of both fields. As a consequence, there is no free reheating parameter. This is particularly important for curvaton-like scenarios, since in these cases the same parameters determine the statistical properties of perturbations and the kinematics of reheating. It is therefore crucial to properly account for the interplay between these two physical effects and the suppression of degeneracies it yields.

When the inflaton has a quadratic potential, extra light scalar fields have recently been studied in \Refs{Bartolo:2002vf,Ellis:2013iea,Byrnes:2014xua,Smith:2015bln} and it has been shown~\cite{Enqvist:2013paa} that the fit of quartic chaotic inflation can be significantly improved in the curvaton limit. In \Ref{Hardwick:2015tma}, a Bayesian analysis was carried out for the quadratic inflaton + curvaton models assuming instantaneous reheating (corresponding to our reheating case number 8), and these models were found not to be disfavoured with respect to standard quadratic inflation.

In this work however, we build a pipeline that incorporates all $\sim 200$ single-field models mapped in \Refs{Martin:2013nzq,Martin:2014vha}, in all $10$ reheating cases. In practice, this means that the Bayesian evidence and complexity of $\sim 2000$ scenarios are addressed, which corresponds to an important step forward in the current state of the art of early Universe Bayesian analysis. 

This paper is organised as follows. In section~\ref{sec:method}, we explain the method that we have employed. We introduce the physical systems under consideration and briefly recall how their predictions were calculated in \Ref{Vennin:2015vfa}. The theory of Bayesian inference is summarised and it is explained how the numerical \ASPIC pipeline~\cite{aspic} was extended to implement scenarios with extra light scalar fields. We then present our results in section~\ref{sec:results}. Reheating cases are analysed one by one for a few prototypical inflaton potentials in section~\ref{sec:PrototypicalPotentials}. Bayesian complexity is introduced as a measure of the number of unconstrained parameters in section~\ref{sec:complexity}, so that the effective number of added parameters compared to single-field setups is quantified. In section~\ref{sec:priorweight}, a procedure of averaging over reheating scenarios is presented, which allows us to derive the Bayesian evidence of categories of models and to discuss the observational status of inflation with an extra light scalar field with respect to standard single-field inflation. Dependence on the prior chosen for the \textit{vev} of the extra light field at the end of inflation is also studied in section~\ref{sec:sigmaendprior}. Finally, in section~\ref{sec:conclusion}, we summarise our results and present a few concluding remarks.
\section{Method}
\label{sec:method}
The method employed in this paper combines the analytical work of \Ref{Vennin:2015vfa} with the numerical tools developed in \Refs{Ringeval:2013lea,Martin:2013nzq}. In this section, we describe its main aspects. 
\subsection{Inflation with an extra light scalar field}
In the present work, we investigate the situation where an extra light scalar field is present during inflation and (p)reheating. The potentials under scrutiny are of the form 
\begin{align}
V(\phi)+\frac{m_\sigma^2}{2}\sigma^2,
\end{align}
where $\sigma$ is taken to be lighter than $\phi$ at the end of inflation. Both fields are assumed to be slowly rolling during inflation, and eventually decay into radiation fluids with decay rates respectively denoted $\Gamma_\phi$ and $\Gamma_\sigma$, during reheating. The parameters describing the inflationary and reheating sectors of the theory are therefore given by
\begin{align}
\theta_{\mathrm{inf}+\mathrm{reh}}=\lbrace \theta_V, m_\sigma, \sigma_\uend,\Gamma_\phi,\Gamma_\sigma\rbrace\, ,
\label{eq:params}
\end{align}
where $\lbrace \theta_V \rbrace$ are the parameters appearing in the inflaton potential $V(\phi)$, $\sigma_\uend$ is the \textit{vev} of $\sigma$ evaluated at the end of inflation, and $m_\sigma$, $\Gamma_\phi$ and $\Gamma_\sigma$ have been defined before. It is important to stress that, as mentioned in the introduction, reheating kinematics is entirely fixed by these parameters, so that the number of \efolds between Hubble exit of the CMB pivot scale and the end of inflation, $\Delta N_*$, only depends on the parameters listed in \Eq{eq:params}.

In \Ref{Vennin:2015vfa}, it is explained how one can make use of the $\delta N$ formalism to relate observables of such models to variations in the energy densities of both fields at the decay time of the last field. This allows us to calculate all relevant physical quantities by only keeping track of the background energy densities. Analytical expressions have been derived for all $10$ reheating cases, that have been implemented in the publicly available \ASPIC library~\cite{aspic}. For a given inflaton potential, and from the input parameters of \Eq{eq:params}, this code returns the value of the three first slow-roll parameters (or equivalently, of the scalar spectral index $\nS$ and its running, and of the tensor-to-scalar ratio $r$) and of the local-type non-Gaussianity parameter $f_{\mathrm{NL}}$.
\subsection{Bayesian Approach to Model Comparison}
\label{sec:bayes}
The next step is to compare the performances of the different inflationary scenarios under consideration. One way to carry out this program is to make use of the Bayesian approach to model comparison~\cite{Cox:1946,Jeffreys:1961,deFinetti:1974,Box:1992,Bernardo:1994,Jaynes:2003,Berger:2003,Trotta:2005ar,Trotta:2008qt}. Bayesian inference uses Bayes theorem to express the posterior probabilities of a set of alternative models $\mathcal{M}_i$ given some data set
$\mathcal{D}$. It reads
\begin{equation}
p\left(\mathcal{M}_i\vert\mathcal{D}\right)
=\frac{\mathcal{E}\left(\mathcal{D}\vert\mathcal{M}_i
\right) \pi\left(\mathcal{M}_i\right)}{p\left(\mathcal{D}\right)}\,
.  
\end{equation} 
Here, $\pi\left(\mathcal{M}_i\right)$ represents the prior belief in
the model $\mathcal{M}_i$, $p\left(\mathcal{D}\right)
=\sum_{i}\mathcal{E}(\mathcal{D}\vert\mathcal{M}_i)\pi(\mathcal{M}_i)$
is a normalisation constant and
$\mathcal{E}\left(\mathcal{D}\vert\mathcal{M}_i \right)$ is the
Bayesian evidence of $\mathcal{M}_i$, defined by
\begin{equation}
\label{eq:evidence:def}
\mathcal{E}\left(\mathcal{D}\vert\mathcal{M}_i \right) 
= \int\dd\theta_{ij}\mathcal{L}
\left(\mathcal{D}\vert\theta_{ij},\mathcal{M}_i\right)
\pi\left(\theta_{ij}\vert \mathcal{M}_i\right)\, ,
\end{equation}
where $\theta_{ij}$ are the $N$ parameters defining the model
$\mathcal{M}_i$ and $\pi\left(\theta_{ij}\vert \mathcal{M}_i\right)$
is their prior distribution. The quantity
$\mathcal{L}\left(\mathcal{D}\vert\theta_{ij},\mathcal{M}_i\right)$, called likelihood function, represents the probability of observing the data $\mathcal{D}$
assuming the model $\mathcal{M}_i$ is true and $\theta_{ij}$ are the
actual values of its parameters. Assuming model $\mathcal{M}_i$, the posterior probability of its parameter $\theta_{ij}$ is then expressed as
\begin{align}
\label{eq:posterior:def}
p\left(\theta_{ij}\vert\mathcal{D},\mathcal{M}_i\right)=\frac{\mathcal{L}
\left(\mathcal{D}\vert\theta_{ij},\mathcal{M}_i\right)
\pi\left(\theta_{ij}\vert \mathcal{M}_i\right)}{\mathcal{E}\left(\mathcal{D}\vert\mathcal{M}_i \right) } \, .
\end{align}

The posterior odds between two models $\mathcal{M}_i$ and
$\mathcal{M}_j$ are given by
\begin{equation}
\frac{p\left(\mathcal{M}_i\vert\mathcal{D}\right)}
{p\left(\mathcal{M}_j\vert\mathcal{D}\right)}
=\frac{\mathcal{E}\left(\mathcal{D}\vert\mathcal{M}_i\right)}
{\mathcal{E}\left(\mathcal{D}\vert\mathcal{M}_j\right)}
\frac{\pi\left(\mathcal{M}_i\right)}{\pi\left(\mathcal{M}_j\right)}\equiv
B_{ij
}\frac{\pi\left(\mathcal{M}_i\right)}{\pi\left(\mathcal{M}_j\right)}\,
, 
\end{equation} 
where we have defined the Bayes factor $B_{ij}$ by
$B_{ij}=\mathcal{E}\left(\mathcal{D}\vert\mathcal{M}_i\right)
/\mathcal{E}\left(\mathcal{D}\vert\mathcal{M}_j\right)$. Under the
principle of indifference, one can assume non-committal model priors,
$\pi(\mathcal{M}_i)=\pi\left(\mathcal{M}_j\right)$, in which case the
Bayes factor becomes identical to the posterior odds (see however section~\ref{sec:priorweight} where another approach is used). With this
assumption, a Bayes factor larger (smaller) than one means a
preference for the model $\mathcal{M}_i$ over the model
$\mathcal{M}_j$ (a preference for $\mathcal{M}_j$ over
$\mathcal{M}_i$). In practice, the ``Jeffreys' scale'' gives an
empirical prescription for translating the values of the Bayes factor
into strengths of belief. When $\ln(B_{ij})>5$, $\mathcal{M}_j$ is
said to be ``strongly disfavoured'' with respect to $\mathcal{M}_i$,
``moderately disfavoured'' if $2.5<\ln(B_{ij})<5$, ``weakly
disfavoured'' if $1<\ln(B_{ij})<2.5$, and the situation is said to be
``inconclusive'' if $\vert \ln(B_{ij})\vert<1$.

Bayesian analysis allows us to identify the models that achieve the
best compromise between quality of the fit and simplicity. In other
words, more complicated descriptions are preferred only if they
provide an improvement in the fit that can compensate for the larger
number of parameters. In the rest of this section, we illustrate how
this idea works in practice on a simple example~\cite{Vennin:2015eaa}.

\begin{figure}[t]
\begin{center}
\includegraphics[width=6cm]{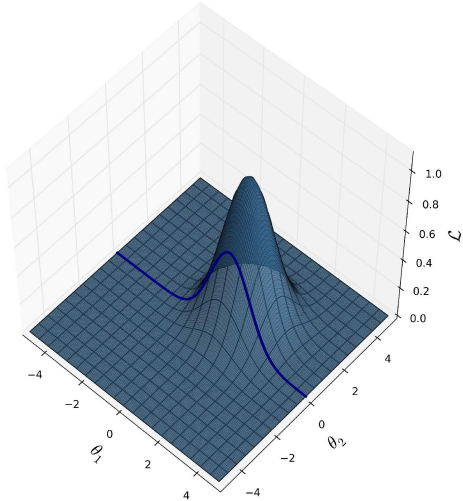}
\caption{Sketch of the likelihood for the toy model $\mathcal{M}_1 $
  discussed in section~\ref{sec:bayes} (pale blue surface). The solid
  blue line corresponds to the likelihood of model $\mathcal{M}_2$,
  which is a sub-model of $\mathcal{M}_1 $ with $\theta_2=0$.}
\label{fig:likelihood}
\end{center}
\end{figure}

Let $\mathcal{M}_1$ and $\mathcal{M}_2$ be two competing models aiming
at explaining some data $\mathcal{D}$. Two parameters $\theta_1$ and
$\theta_2$ describe the first model $\mathcal{M}_1$, and we assume
that the likelihood function is a Gaussian centred at
$(\bar{\theta}_1,\bar{\theta}_2)$ with standard deviations $\sigma_1$
and $\sigma_2$,
\begin{equation}
\label{eq:likelihood:M1}
\mathcal{L}\left(\mathcal{D}\vert\theta_{1},\theta_2,\mathcal{M}_1\right)=
 \mathcal{L}_1^\mathrm{max} 
e^{-\frac{\left(\theta_1-\bar{\theta}_1\right)^2}{2\sigma_1^2}
-\frac{\left(\theta_2-\bar{\theta}_2\right)^2}{2\sigma_2^2}} .
\end{equation}
This likelihood is represented in \Fig{fig:likelihood}. We assume that
the prior distribution on $\theta_1$ and $\theta_2$ is also a
Gaussian, with standard deviations $\Sigma_1$ and $\Sigma_2$, and that
the likelihood is much more peaked than the prior, that is to say
$\Sigma_1\gg \sigma_1$ and $\Sigma_2\gg \sigma_2$. In this limit,
\Eq{eq:evidence:def} gives rise to a simple expression for the
evidence of model $\mathcal{M}_1$, namely
\begin{equation}
\label{eq:evid:M1}
\mathcal{E}\left(\mathcal{D}\vert\mathcal{M}_1 \right) 
= \frac{\sigma_1\sigma_2}{\Sigma_1\Sigma_2} \mathcal{L}_1^\mathrm{max}\, .
\end{equation}
One can readily see that the higher the best fit,
$\mathcal{L}_1^\mathrm{max}$, the better the Bayesian evidence, which
is of course expected. On the other hand, the ratio
$\sigma_1\sigma_2/(\Sigma_1\Sigma_2)$ stands for the volume reduction
in parameter space induced by the data $\mathcal{D}$ and, therefore,
quantifies how much the parameters $\theta_1$ and $\theta_2$ must be
fine tuned around the preferred values $\bar{\theta}_1$ and
$\bar{\theta}_2$ to account for the data. From \Eq{eq:evid:M1}, it is
thus clear that the larger this fine tuning, the worse the Bayesian
evidence, the final result being a trade-off between both effects.

Now, let us imagine that the second parameter $\theta_2$ is associated
with some extra, non-minimal, feature (such as, say, isocurvature
perturbations, non-Gaussianities, oscillations in the power
spectrum, \textit{etc}). We want to determine whether $\theta_2$ (and, hence, the
associated feature) is, at a statistically significant level, required
by the data. To this end, we introduce the model $\mathcal{M}_2$ that
is a sub-model of $\mathcal{M}_1$ where we choose $\theta_2=0$. This
new model $\mathcal{M}_2$ has a single parameter $\theta_1$. By
definition, its prior distribution is Gaussian with standard deviation
$\Sigma_1$ and its likelihood function is given by
\begin{eqnarray}
\mathcal{L}\left(\mathcal{D}\vert\theta_{1},\mathcal{M}_2\right)&=&
\mathcal{L}\left(\mathcal{D}\vert\theta_{1},0,\mathcal{M}_1\right)
=\mathcal{L}_1^\mathrm{max}e^{-\frac{\bar{\theta}_2^2}{2\sigma_2^2}} 
e^{-\frac{\left(\theta_1-\bar{\theta}_1\right)^2}{2\sigma_1^2}} 
\equiv \mathcal{L}_2^\mathrm{max}
e^{-\frac{\left(\theta_1-\bar{\theta}_1\right)^2}{2\sigma_1^2}}\, ,
\end{eqnarray} 
where we have defined the maximum likelihood for model $\mathcal{M}_2$ by
$\mathcal{L}_2^\mathrm{max}=\mathcal{L}_1^\mathrm{max}
\exp[-\bar{\theta}_2^2/(2\sigma_2^2)]$. This likelihood is displayed
as the solid blue line in \Fig{fig:likelihood}, and simply corresponds
to the intersection of the full likelihood~(\ref{eq:likelihood:M1})
with the plane $\theta_2=0$. In the same limit $\Sigma_1\gg \sigma_1$
as before, the evidence for the model $\mathcal{M}_2$ is given by an
expression similar to \Eq{eq:evid:M1}, namely
$\mathcal{E}(\mathcal{D}\vert\mathcal{M}_2) = \sigma_1
\mathcal{L}_2^\mathrm{max}/\Sigma_1$. The Bayes factor between models
$\mathcal{M}_1$ and $\mathcal{M}_2$ therefore reads
\begin{equation}
B_{12}=\frac{\mathcal{L}_1^\mathrm{max}}{\mathcal{L}_2^\mathrm{max}}
\frac{\sigma_2}{\Sigma_2}
= e^{\frac{\bar{\theta}_2^2}{2\sigma_2^2}}\frac{\sigma_2}{\Sigma_2}\,.  
\end{equation} 
The first term,
$\mathcal{L}_1^\mathrm{max}/\mathcal{L}_2^\mathrm{max}$ represents the
change in the best fit due to the fact
that we have added new parameters. Obviously, this ratio is
always larger than one since adding more degrees of freedom to
describe the data can only improve the quality of the fit. On the
other hand, the second term $\sigma_2/\Sigma_2$ represents the amount
of fine tuning required for this new parameter $\theta_2$ and is
smaller than one. As a consequence, if the improvement of the fit
quality is not large enough to beat fine tuning, one concludes that
the parameter $\theta_2$ is not required by the data. In the opposite
case, one concludes that there is a statistically significant
indication that $\theta_2\neq 0$. Finally, let us notice that if the data is completely insensitive to $\theta_2$, $\mathcal{L}_1^\mathrm{max}=\mathcal{L}_2^\mathrm{max}$ and $\sigma_2=\Sigma_2$, the two models $\mathcal{M}_1$ and $\mathcal{M}_2$ have the same Bayesian evidence. Bayesian evidence is therefore
insensitive to unconstrained parameters (in such a case, $\mathcal{M}_1$ and $\mathcal{M}_2$ can be differentiated using Bayesian complexity, see section~\ref{sec:complexity}).
\subsection{Fast Bayesian Evidence Computation}
The computation of Bayesian evidence is a numerically expensive task, since it requires one to evaluate the multi-dimensional integral of \Eq{eq:evidence:def}. A typical analysis based on the Planck likelihood coupled with an exact inflationary code to integrate the perturbations typically requires more than 3 CPU years of computing time on standard modern processors. Given the large number of models over which we want to carry out the Bayesian programme, this means that these conventional methods cannot be employed in a reasonable amount of time. This is why we resort to two important simplifications.

First, perturbations are calculated making use of the slow-roll formalism during inflation, and the $\delta N$ formalism (for curvature perturbations) afterwards. These two approaches provide analytical expressions for the power spectra of scalar curvature and tensor fluctuations, and for the local non-Gaussianity level. They exempt us from using a numerical integrator of the mode equations.

Second, we use the ``effective likelihood via slow-roll reparametrisation'' approach proposed by Christophe Ringeval in \Ref{Ringeval:2013lea}. The method relies on the determination of an effective likelihood for inflation, which is a function of the primordial amplitude of the scalar perturbations complemented with the necessary number of slow-roll parameters to reach the desired accuracy. The effective likelihood is obtained by marginalisation over the standard cosmological parameters, viewed as ``nuisance'' from the early Universe point of view. Machine-learning algorithms are then used to reproduce the multidimensional shape of the likelihood, and Bayesian inference is carried out with the nested sampling algorithm \MULTINEST. The high accuracy of the method, which increases by orders of magnitude the speed of performing Bayesian inference and parameter estimation in an inflationary context, has been confirmed in \Ref{Ringeval:2013lea}. We use this procedure~\cite{Ringeval:PC} with the Planck 2015 $TT$ data combined with high-$\ell$ $C_\ell^{TE}+C_\ell^{EE}$ likelihood and low-$\ell$ temperature plus polarization likelihood (PlanckTT,TE,EE+lowTEB in the notations of \Ref{Aghanim:2015xee}, see Fig.~1 there), together with the BICEP2-Keck/Planck likelihood described in \Ref{Ade:2015tva}.
\subsection{Prior Choices}
\label{sec:priors}
The priors encode physical information one has a priori on the values of the parameters~(\ref{eq:params}) that describe the models. For the parameters of the potential $\lbrace \theta_V \rbrace$, we use the same priors as the ones proposed in \Ref{Martin:2013nzq}, which are based on the physical, model-building related, considerations of \Ref{Martin:2014vha}. Because the extra field $\sigma$ is supposed to be still light at the end of inflation, its mass $m_\sigma$ must be smaller than the Hubble scale at the end of inflation, $H_\uend$. The same condition applies to the two decay rates, $\Gamma_\phi,\ \Gamma_\sigma<H_\uend$, since both fields decay after inflation. On the other hand, we want the Universe to have fully reheated before Big Bang Nucleosynthesis (BBN), which means that the two decay rates are also bounded from below by $H_{\mathrm{BBN}}\simeq (10\MeV)^2/\Mp$. The same lower bound applies to $m_\sigma$ since, assuming perturbative decay, $m_\sigma>\Gamma_\sigma$. Between these two values, the order of magnitude of $m_\sigma$ and of the two decay rates is unknown, which is why a logarithmically flat prior (or ``Jeffreys prior'') is chosen:
\begin{align}
\ln H_{\mathrm{BBN}} < \ln \Gamma_\phi,\,\ln\Gamma_\sigma,\,\ln m_\sigma < \ln H_\uend\, ,
\label{eq:prior:massscales}
\end{align}
where, for a given reheating case, the extra-ordering conditions given in \Fig{fig:rho} are further imposed. 

Two kinds of priors are then considered for $\sigma_\uend$. A first approach corresponds to stating that the order of magnitude of $\sigma_\uend$ is unknown, and that a logarithmically flat prior on $\sigma_\uend$ should be employed
\begin{align}
\label{eq:sigmaend:LogPrior}
\ln\sigma_\uend^\umin < \ln\sigma_\uend < \ln \sigma_\uend^\umax \, .
\end{align}
Here, $\sigma_\uend^\umin$ and $\sigma_\uend^\umax$ are the boundary values given for each reheating case in \Fig{fig:rho}. A second approach relies on the equilibrium distribution\footnote{In \Ref{Enqvist:2012}, it is shown that the timescale of equilibration depends on $m_\sigma$ in practice, but can be surprisingly large (even more than thousands of $e$-folds). This implies that the initial conditions for spectator fields are not automatically erased during inflation, and that time variation of $H$ in (even slow-roll) inflation could play a role as well. Therefore, distributions different from \Eq{eq:sigmaend:GaussianPrior} may also be relevant and this is another reason why, in this work, only the rough condition $m_\sigma\sigma_\uend\sim H_\uend^2$ is imposed.} of long wavelength modes of a light spectator field in de Sitter~\cite{Starobinsky:1986fx, Enqvist:2012}
\begin{align}
P\left(\sigma_\uend\right) \propto \exp\left(-\frac{4\pi^2 m_\sigma^2\sigma_\uend^2}{3H_\uend^4}\right)\, .
\label{eq:sigmaend:GaussianPrior}
\end{align}
This distribution is often referred to as the ``Gaussian prior'' for $\sigma_\uend$. However, one should note that the Hubble scale at the end of inflation, $H_\uend$, that appears in \Eq{eq:sigmaend:GaussianPrior}, is in fact a function of $\sigma_\uend$. Indeed, it depends on the mass scale of the inflaton potential, which is fixed to reproduce a given value of the curvature primordial power spectrum amplitude $P_*$. Since $\sigma$ contributes to the total amount of scalar perturbations, $P_*$ explicitly depends on $\sigma_\uend$, and so does $H_\uend$. This is why, in \Eq{eq:sigmaend:GaussianPrior}, one should write $H_\uend(\sigma_\uend)$ and the corresponding distribution is not, strictly speaking, a Gaussian. 
Its physical interpretation should therefore be handled carefully. In this work, we thus take a more pragmatic approach and interpret \Eq{eq:sigmaend:GaussianPrior} only as the idea that physically acceptable values of the parameters should be such that $m_\sigma\sigma_\uend\sim H_\uend^2$, in agreement with what one would expect for a light scalar field in de Sitter. In practice, we implement this requirement by simply rejecting realisations for which the argument of the exponential function in \Eq{eq:sigmaend:GaussianPrior} is smaller than $1/10$ or larger than $10$ (we have checked that when changing these arbitrary values to, say, $1/100$ and $100$, very similar results are obtained).
\section{Results}
\label{sec:results}
In this section we cast our results in a series of a few tables. Let us first explain which quantities are displayed. In the following tables, the first column is an acronym for the name of the inflationary scenario under consideration that follows the same conventions as in \Ref{Vennin:2015vfa}. The index appearing after ``MC'' refers to the reheating case number, while the second part of the acronym stands for the name of the inflaton potential, following \Ref{Martin:2014vha}. For example, $\mathrm{MC}_3\mathrm{LFI}_2$ corresponds to the case where the inflaton potential is of the large field, quadratic type, and where the reheating scenario is of the third kind (see \Fig{fig:rho}). In the second column is given the logarithm of the Bayesian evidence, defined in section~\ref{sec:bayes}, normalised to the Bayesian evidence of (the single-field version of) Higgs Inflation (the Starobinsky model). Normalisation choice is anyway arbitrary and what only makes physical sense is ratios of Bayesian evidence, but we chose Higgs Inflation to be the reference model in order to match the convention of \Ref{Martin:2013nzq} and to make comparison with this work easier. The third and fourth columns respectively stand for the number of input parameters and the number of unconstrained parameters, that will be defined and commented on in section~\ref{sec:complexity}. Finally, the fifth and last column gives the maximal value of the likelihood (``best fit''), still normalised to the Bayesian evidence of Higgs Inflation. This quantity is irrelevant from a purely Bayesian perspective (it would only need to be considered in a frequentist analysis), but we display it for indicative purpose and as it allows one to check consistency with the results presented in \Ref{Vennin:2015vfa} where exploration in parameter space in performed.

For comparison, in all tables displayed below, the first line corresponds to the single-field version of the models under consideration, in the case where the mean equation of state parameter during reheating vanishes, $\bar{w}_\ureh=0$, and the energy density at the end of reheating, $\rho_\ureh$, has the same logarithmically flat prior as in \Eq{eq:prior:massscales}, $\ln\rho_{{}_\mathrm{BBN}}<\ln\rho_\ureh<\ln\rho_\uend$. The reason is that, in this work, the inflaton is assumed to oscillate around a quadratic minimum of its potential after inflation ends (in which case its energy density redshits as matter), and we want the limit where $\sigma_\uend\rightarrow 0$ to match the single-field version of the model (even if subtleties regarding this limit are to be noted, see section~\ref{sec:PrototypicalPotentials}). Finally, let us mention that apart from section~\ref{sec:sigmaendprior}, the results given here are obtained from the logarithmically flat prior~(\ref{eq:sigmaend:LogPrior}) on $\sigma_\uend$.
\subsection{Prototypical Inflaton Potentials}
\label{sec:PrototypicalPotentials}
As already mentioned, the inclusion of an extra light scalar field in the \ASPIC pipeline, for all ten reheating scenarios, gives rise to $\sim 2000$ models of inflation for which the Bayesian programme can be carried out. In this paper, for conciseness, we choose not to display all corresponding Bayesian evidence and select four prototypical inflaton potentials that will allow us to discuss the main generic trends that we have more generically observed. These four examples are also discussed in great detail in \Ref{Vennin:2015vfa}, where, in all ten reheating cases, plots in the $(\nS,r)$, $(\nS,f_{_\mathrm{NL}})$ and $(f_{_\mathrm{NL}},r)$ planes are provided. Together with the table 1 of this same reference where the main properties of these figures are summarised, they constitute useful prerequisites to properly interpret the following results.
\begin{itemize}
\item Large-field inflation (LFI) is a typical example of a ``large-field'' model. Its potential is given by
\begin{align}
\label{eq:lfi:pot}
V\left(\phi\right) = M^4\left(\frac{\phi}{\Mp}\right)^p\, .
\end{align}
Here, $p$ is the free parameter of the potential. In this section, we present the results obtained for $p=2$ (i.e. for a quadratic potential) but the results obtained for $p=2/3$, $p=1$, $p=3$ and $p=4$, as well as for marginalising over $p\in [0.2,5]$, are given in appendix~\ref{sec:lfi:otherp} (and in appendix~\ref{sec:Gaussian} for a Gaussian prior on $\sigma_\uend$). Large-field models are well-known for yielding a value for $r$ that is too large in their single-field versions. Since the introduction of light scalar fields typically reduces the predicted value of $r$, at least in reheating cases 4, 5, 7 and 8, we expect the Bayesian evidence of these models to be modified.
\item Higgs inflation (HI, the Starobinsky model) is a typical example of a ``Plateau model'' for which the single-field version of the model already provides a very good fit to the data (its Bayesian evidence can therefore only decrease). Its potential is given by
\begin{align}
V\left(\phi\right) = M^4 \left[1-\exp\left(-\sqrt{\frac{2}{3}}\frac{\phi}{\Mp}\right)\right]^2\, .
\end{align}
\item  Natural Inflation (NI) has a potential given by
\begin{align}
V\left(\phi\right) = M^4\left[1+\cos\left(\frac{\phi}{f}\right)\right]\, .
\end{align}
When $f$ is not super-Planckian, it is a typical example that yields a value for $\nS$ that is too small in the single-field version of the model. The introduction of a light scalar field tends to drive $\nS$ towards $1$ so that the predictions of the models intersect the region that is preferred by the data, but only for fine-tuned parameters. How the Bayesian evidence will change is therefore difficult to predict. As in \Ref{Martin:2013nzq}, a logarithmically flat prior is chosen on $f$, $0<\log(f/\Mp)<2.5$.
\item Power-Law Inflation (PLI) is a typical potential yielding a too large value of $\nS$, or a too large value of $r$. Since extra light scalar fields tend to decrease $r$ but to increase $\nS$, it is also difficult to predict how the Bayesian evidence will evolve. Its potential is given by
\begin{align}
V\left(\phi\right) = M^4\exp\left(-\alpha\frac{\phi}{\Mp}\right)\, .
\end{align}
Since this potential has the specific feature to be conformally invariant, its predictions do not depend on the number of \efolds $\Delta N_*$ elapsed between Hubble exit and the end of inflation, hence do not depend on the reheating kinematics. This is another reason why this example is interesting, since it isolates the effects of $\sigma$ coming from its contribution to the total amount of scalar perturbations, and is not sensitive to the role it plays in the reheating dynamics. As in \Ref{Martin:2013nzq}, a logarithmically flat prior is chosen on $\alpha$, $-4<\log\alpha<-1$.
\end{itemize}

The results for Higgs Inflation and Natural Inflation are presented in table~\ref{table:HINI} and the results for Large-Field (quadratic) Inflation and Power-Law Inflation are given in table~\ref{table:LFI2PLI}.

\tableHINI

A first remark concerns the single-field limit. The first reheating scenario (see \Fig{fig:rho}) corresponds to values of the parameters such that the extra light field never dominates the energy budget of the Universe, and decays before the inflaton field does. In this case, it does not contribute to curvature perturbations neither does it play a role in the reheating dynamics. As a consequence, the first reheating scenario can be viewed as the single-field limit of the models under consideration, and in \Ref{Vennin:2015vfa}, it is shown that the same predictions as for the single-field models are indeed obtained in this case. However, in tables~\ref{table:HINI} and~\ref{table:LFI2PLI}, one can see that the Bayesian evidence for the first reheating scenarios (``$\mathrm{MC}_1$'') and the single-field models are not exactly the same. The reason is that reheating is effectively implemented with different priors. In the single-field models, as explained in the beginning of section~\ref{sec:results}, the total energy density at the end of reheating is drawn according to a logarithmically flat prior $\ln \rho_{{}_\mathrm{BBN}}<\ln \rho_\ureh< \ln \rho_\uend$. In the first reheating scenario on the other hand, $\rho_\ureh\simeq 3\Mp^2\Gamma_\phi^2$ which is drawn according to the same distribution, with the difference that the condition $\Gamma_\phi < \Gamma_\sigma$ is further implemented. This means that, when marginalised over all other parameters, the effective prior distribution for $\Gamma_\phi$ alone is biased towards smaller values (for which more values of $\Gamma_\sigma$ are allowed). In the first reheating scenario, $\bar{w}_\ureh\simeq 0$, which means that~\cite{Martin:2010kz,Easther:2011yq} $\Delta N_* \propto \ln \Gamma_\phi$ shows preference for smaller values too. As a consequence, steeper portions of the inflaton potential are preferentially sampled in the first reheating scenario, hence slightly lower Bayesian evidence are obtained (an important exception being Power-Law Inflation which is, as explained above, insensitive to reheating kinematic effects). This again shows the importance of properly accounting for reheating kinematic effects. The second reheating scenario (``$\mathrm{MC}_2$'') is similar in the sense that the extra light field decays before the inflaton does, and is subdominant when this happens. In \Ref{Vennin:2015vfa}, it is shown that almost the same predictions as for the single-field models are obtained in this case too. However, reheating dynamics is more complicated and contrary to case $1$ where reheating is only made of a matter phase, case $2$ contains a matter phase, a second inflation phase, an other matter phase, a radiation phase and then a matter phase again. In practice, it turns out that the values of $\Delta N_*$ it yields are closer to the single-field ones, which is why the effect described here is smaller in these cases.

\tableLFITWOPLI

For the other reheating scenarios, predictions can be very different from the single-field ones, and we now review our four prototypical potentials one by one. For Higgs Inflation (table~\ref{table:HINI}, left panel), since the single-field version of the model already provides one of the best possible fits to the data, as expected, including an extra light scalar field leads to decreasing the Bayesian evidence in all reheating scenarios. However, according to the Jeffrey scale, this change is ``inconclusive'' in most cases (and only ``moderately disfavoured'' in reheating scenarios $3$ and $10$). This means that, at the Bayesian level, Plateau models such as Higgs Inflation are robust under the introduction of extra light degrees of freedom. In reheating scenarios $4$, $5$, $7$ and $8$, $\nS$ varies between the single-field prediction and $1$, which explains why the evidence decreases. The main differences between these cases is that in scenarios $4$ and $7$, large non-Gaussianities can also be produced while in scenarios $5$ and $8$, non-Gaussianities always remain within the observational bounds. For this reason, it may seem counter-intuitive that the evidence of cases $5$ and $8$ is slightly lower than the one of cases $4$ and $7$. This means that non-Gaussianities play a negligible role in constraining these models, and that most parameters that are excluded because of too large non-Gaussianities are already excluded because of too large values of $\nS$. Non-Gaussianities observational constrains are therefore still too weak to really constrain these models. Finally, reheating scenarios $3$, $6$, $9$ and $10$ contain a second phase of inflation which means that the scales observed in the CMB exit the Hubble radius closer to the end of inflation where the potential is steeper (hence smaller values of $\nS$ and larger values for $r$), which is why these models have lower evidence, the most disfavoured scenario being case $10$.

For Natural Inflation (table~\ref{table:HINI}, right panel), let us again note that no radical change is observed when introducing a light scalar field. The single-field version of the model produces a too small value for $\nS$ which is why it is ``moderately disfavoured'' compared to, say, Higgs Inflation. In reheating scenarios $4$, $5$, $7$ and $8$, $\nS$ varies continuously between the single-field prediction and $1$, crossing the ``sweet spot'' of the data. However, this only yields a small increase of the Bayesian evidence of these four cases compared to reheating scenario $\mathrm{MC}_1$, confirming that, as already mentioned, parameters achieving the right scalar tilt are fine-tuned. In fact, only cases $5$ and $8$ are preferred to the single-field version of the model (although at an ``inconclusive'' level). Indeed, only in these cases, non-Gaussianities remain within observational constrains. This means that contrary to Higgs Inflation, current observational constraints on non-Gaussianities are already relevant for these models. Finally, cases $3$, $6$, $9$ and $10$ produce values of $\nS$ that are even smaller than the single-field predictions, and are therefore disfavoured. The worst scenario is again $10$, but it is only weakly disfavoured with respect to the single-field version of the model.

For large-field models such as Large-Field Inflation, depending on the power index $p$ of the potential~(\ref{eq:lfi:pot}), larger modifications in the Bayesian evidence can be observed. The values for a quadratic potential are displayed in the left panel of table~\ref{table:LFI2PLI} while other values of $p$ are dealt with in appendix~\ref{sec:lfi:otherp}. The single-field versions of the large-field models suffer from a value for $r$ which is too large compared with the current observational constrains. In scenarios $4$, $5$, $7$ and $8$, the corresponding value decreases. This is why, in cases $5$ and $8$ where no large non-Gaussianities are produced, one obtains larger Bayesian evidence than in the single-field case. For a quadratic potential, the difference in only ``inconclusive'' (with respect to single-field) or ``weak'' (with respect to $\mathrm{MC}_1$) according to the Jeffreys scale, and the best scenario ($\mclfiFIVETWO$) still remains ``moderately disfavoured'' with respect to the best single-field models such as Higgs inflation. For a quartic potential however (see appendix~\ref{sec:lfi:otherp}), the effect is much larger as cases $5$ and $8$ are ``strongly favoured'' with respect to their single-field counterpart. The best scenario, $\mclfiFIVEFOUR$, is even in the ``inconclusive'' zone of Higgs inflation. From a Bayesian perspective, this scenario therefore belongs to the best models of inflation. On the contrary, because large non-Gaussianities are produced in cases $4$ and $7$, the decrease in $r$ is not enough to ``rescue'' these models that even have slightly worse Bayesian evidence than their single-field versions. This means that, as in Natural Inflation but contrary to Higgs Inflation, non-Gaussianities already are a constraining observable for these models. In cases $3$, $6$, $9$ and $10$, smaller values of $r$ are obtained when $p>2$, but $\nS$ reaches unacceptably small values in these regimes. This is why these scenarios are disfavoured, even compared to their single-field counterpart.

Finally, Power-Law Inflation is displayed in the right panel of table~\ref{table:LFI2PLI}. Being conformally invariant, its predictions do not depend on $\Delta N_*$, and there is no reheating kinematic effects in these models. This is why, as already noticed, the reheating scenarios $\mathrm{MC}_1$ and $\mathrm{MC}_2$ have exactly the same Bayesian evidence as the single-field version of the model. In practice, the Bayesian evidence of all reheating scenarios are extremely low and close to the numerical limit of our code, so it is not easy to well resolve the differences between them. However, one can see that all scenarios are within the same ``inconclusive'' zone. This means that they are all equally strongly disfavoured. This also reinforces the statement that, for the models considered in the present paper in particular and for inflation in general, reheating kinematic effects play an important role, to which the data are now sensitive.
\subsection{Complexity and Number of Unconstrained Parameters}
\label{sec:complexity}
As shown in section~\ref{sec:bayes}, Bayesian evidence is not sensitive to unconstrained parameters. Concretely, this means that if one adds a new parameter to a model, if this parameter does not change any of the model predictions, the Bayesian evidence remains the same. In this case, the model with less parameters may be considered more ``simple''. This is why the idea of ``complexity'' naturally arises in Bayesian analysis. At first sight, the models considered here are more ``complex'' than pure single-field scenarios as they contain more parameters. However, the data may be more efficient in constraining these parameters and the relevant question is rather how many unconstrained parameters these models have. A well-suited measure of the number of free parameters that the data can actually constrain in a model is provided by the relative entropy between the prior and posterior distributions~\cite{Spiegelhalter} (the Kullback-Leibler divergence). In \Ref{Kunz:2006mc}, it is shown that such an effective number of parameters, called Bayesian complexity $\mathcal{C}$, can be written as
\begin{align}
\mathcal{C}_i = -2 \left\langle \ln\mathcal{L}\left(\theta_{ij}\right) \right\rangle_p + 2 \ln \mathcal{L}\left(\theta_{ij}^\mathrm{ML}\right)\, ,
\end{align}
where $\langle\cdot \rangle_p$ denotes averaging over the posterior $p\left(\theta_{ij}\vert\mathcal{D},\mathcal{M}_i\right)$ and $\theta_{ij}^\mathrm{ML}$ is the parameters values where the likelihood is maximal. Bayesian complexity therefore assesses the constraining power of the data with respect to the measure provided by the prior. 
\begin{figure}[t]
\begin{center}
\includegraphics[width=6cm]{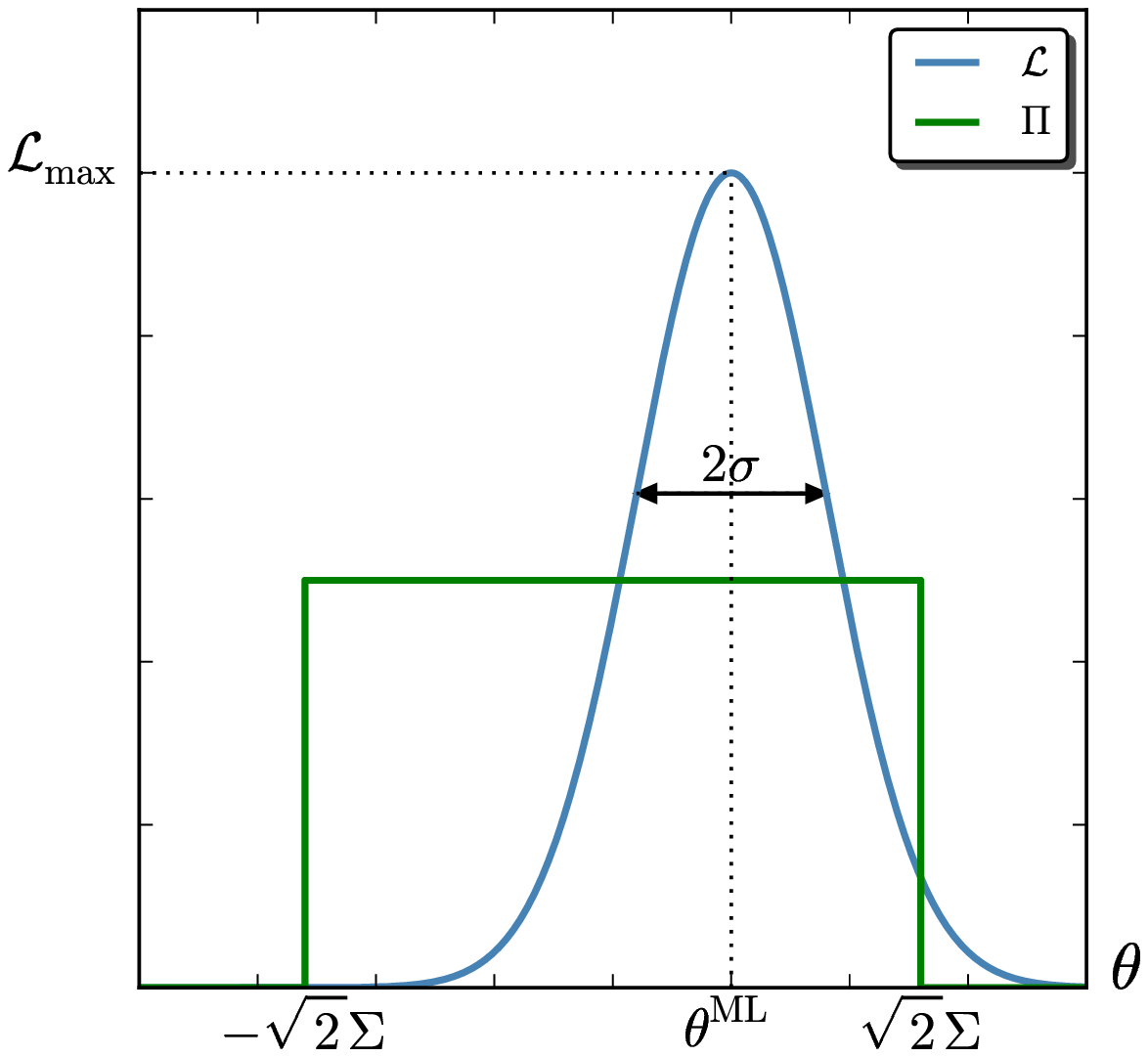}
\includegraphics[width=6cm]{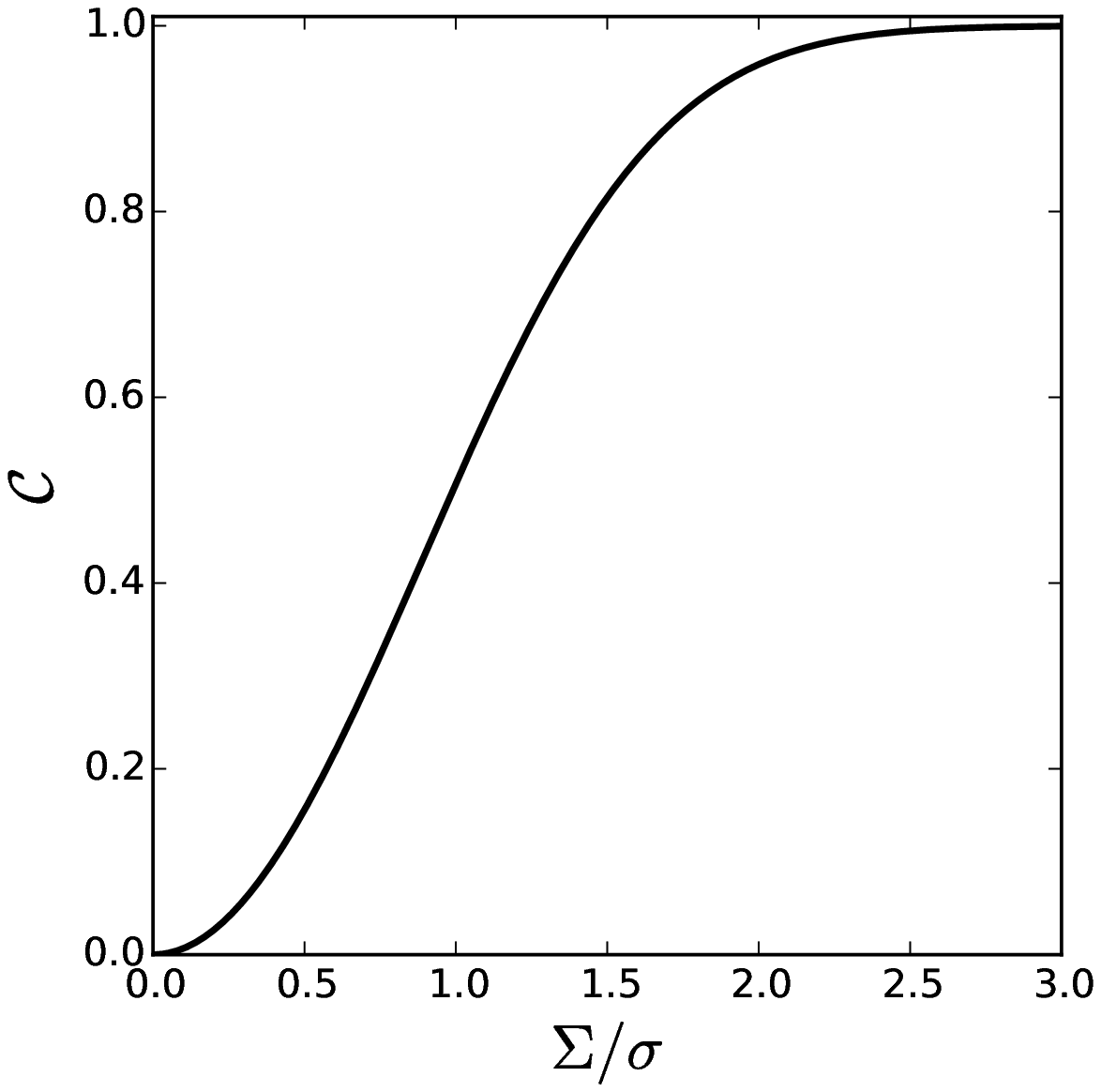}
\caption{Toy model described in section~\ref{sec:complexity}. In the left panel, the flat prior $\Pi$ with standard deviation $\Sigma$, and the Gaussian likelihood $\mathcal{L}$ that peaks at $\theta^\mathrm{ML}$ and has standard deviation $\sigma$, are displayed. In the right panel, the Bayesian complexity $\mathcal{C}$ of the model is given in terms of $\Sigma/\sigma$.}
\label{fig:complexity:toymodel}
\end{center}
\end{figure}

To see how this works in practice, let us consider the toy example depicted in the left panel of \Fig{fig:complexity:toymodel}. This model has a single parameter $\theta$ with a flat prior between $-\sqrt{2}\Sigma$ and $\sqrt{2}\Sigma$ ($\Sigma$ is the prior standard deviation). Let us assume the likelihood to be Gaussian, centred over $\theta^\mathrm{ML}$ and with standard deviation $\sigma$. For simplicity, in the expressions given below, let us assume that $\theta^\mathrm{ML}=0$ so that the prior and likelihood are centred over the same value. Making use of \Eq{eq:evidence:def}, the Bayesian evidence of this model is given by
\begin{align}
\label{eq:evidence:toymodel:complexity}
\mathcal{E}=\frac{\sqrt{\pi}}{2}\mathcal{L}_\mathrm{max} \frac{\sigma}{\Sigma}\erf\left(\frac{\Sigma}{\sigma}\right)\, ,
\end{align}
and one recovers the features commented on in section~\ref{sec:bayes}, namely the fact that the Bayesian evidence increases with $\mathcal{L}_\mathrm{max}$ but decreases with $\Sigma/\sigma$. Using \Eq{eq:posterior:def}, the posterior distribution of the model can then be calculated, and one obtains for the Bayesian complexity
\begin{align}
\mathcal{C}= 1-\frac{2}{\sqrt{\pi}}\frac{\Sigma}{\sigma}\frac{\ee^{-\Sigma^2/\sigma^2}}{\erf\left(\Sigma/\sigma\right)}\, .
\end{align}
One can see that, contrary to the Bayesian evidence in \Eq{eq:evidence:toymodel:complexity}, the complexity only depends on the ratio $\Sigma/\sigma$ and not on $\mathcal{L}_\mathrm{max}$. It is displayed in the right panel of \Fig{fig:complexity:toymodel}. When $\Sigma\gg \sigma$, $\theta$ is well measured and accordingly, the complexity (i.e. the number of constrained parameters) goes to one. On the contrary, when $\Sigma\ll\sigma$, the data does not constrain the parameter $\theta$ and the complexity vanishes.

From this example, it is clear that Bayesian complexity allows us to quantify the number of unconstrained parameters
\begin{align}
\UnconstrainedParams = N - \mathcal{C}\, .
\label{eq:unconParam}
\end{align}
The models considered in the present work have the same number of inflaton potential parameters as their single-field counterpart. As already mentioned, see \Eq{eq:params}, their predictions also depend on $\lbrace m_\sigma,\sigma_\uend,\Gamma_\phi,\Gamma_\sigma\rbrace$, while their single-field analogues only rely on $\rho_\mathrm{reh}$ to describe the reheating sector. This is why $3$ more parameters are involved, which can be checked in the third column of the tables displayed above. However, the number of unconstrained parameters differs in general by a different amount when comparing single-field models to cases where an extra light field is present.
In practice, one can check that this amount is always larger than $0$ (except for Power-Law potentials, but see below), hence there are more unconstrained parameters, but also always smaller than $3$, hence more parameters are constrained. For $\mclfiFIVETWO$ for example, {\pgfmathparse{3+\NUPlfiTWO-\NUPmclfiFIVETWO}\pgfmathprintnumber{\pgfmathresult}} more parameters are constrained with respect to the single-field version of the model. This means that the data is more efficient in constraining parameters when an extra light scalar field is added. This should be related to the fact that, as already stressed, the added parameters not only contribute to the reheating kinematic description (which, at the effective level, boils down to one single parameter), but also to the statistical properties of the perturbations themselves to which the added field contributes. The interplay between these two effects lead to a suppression of degeneracies that Bayesian complexity is therefore quantifying.

Finally, let us notice that for Power-Law Inflation, as well as for some reheating scenarios of Natural Inflation, negative numbers of unconstrained parameters can be obtained. This is because, when the best-fit likelihood of a model is very poor, its Bayesian complexity can be arbitrarily large, hence $\UnconstrainedParams$ in \Eq{eq:unconParam} becomes negative. In some sense, it means that the model is so strongly disfavoured that the data calls for more parameters than it can offer.

\subsection{Averaging over Reheating Scenarios}
\label{sec:priorweight}
\tableHINIAverage
So far, Bayesian evidence of inflationary models where an extra light scalar field is present have been given for each reheating scenario individually. However, in order to determine whether predictions of purely single-field models of inflation are robust under the introduction of light scalar fields, and whether the inflaton potentials for which the data show the strongest preference change once these extra fields are accounted for, one may want to derive ``consolidated'' Bayesian evidence associated to the inflaton potential only, regardless of the reheating scenario. This can be done by averaging over reheating scenarios in the following manner.
For the purpose of illustration, let us consider two toy models $\mathcal{M}_1$ and $\mathcal{M}_2$, that both depend on the same parameter $\theta$. In model $\mathcal{M}_1$, $\theta$ is assumed to lie within the range $[a,b]$ with a flat prior distribution, while in model $\mathcal{M}_2$, $\theta$ lies within the range $[b,c]$ with a flat prior distribution too. The model $\mathcal{M}_{1+2}$ is defined to be the ``union'' of $\mathcal{M}_1$ and $\mathcal{M}_2$, where $\theta$ lies in $[a,c]$ with a flat prior distribution, so that $\mathcal{M}_1$ and $\mathcal{M}_2$ are simply sub-models of $\mathcal{M}_{1+2}$. From the definition~(\ref{eq:evidence:def}), one readily obtains
\begin{align}
\mathcal{E}\left(\mathcal{D}\vert \mathcal{M}_{1+2}\right) = \frac{b-a}{c-a}\mathcal{E}\left(\mathcal{D}\vert \mathcal{M}_{1}\right) + \frac{c-b}{c-a}\mathcal{E}\left(\mathcal{D}\vert \mathcal{M}_{2}\right)\, .
\end{align}
In other words, the Bayesian evidence of model $\mathcal{M}_{1+2}$ is obtained by averaging the evidence of models $\mathcal{M}_1$ and $\mathcal{M}_2$, weighted by the relative fraction of the prior volume of $\mathcal{M}_{1+2}$ that falls into their respective domains. Somehow, these prior volume fractions can be viewed as priors for the sub-models $\mathcal{M}_1$ and $\mathcal{M}_2$ themselves.

This is the strategy we adopt here. Notice that it requires the prior spaces associated with the different reheating scenarios to be disjoint and one can check that, from \Fig{fig:rho}, it is indeed the case. In practice, starting from the global priors~(\ref{eq:prior:massscales}), and using a fiducial, constant likelihood in our Bayesian inference code, we compute the fraction of attempts that fall into each of the ten reheating scenarios. The corresponding ten relative weights are given in the second columns of tables~\ref{table:HINI:Average} and~\ref{table:LFI2PLI:Average} for the four potentials considered in this section (the analysis for other monomial large-field potentials is presented in appendix~\ref{sec:largefield:average}). One can see that these weights are roughly independent of the inflaton potential, small differences being due to hard prior conditions (notably the requirement that the scalar power spectrum can properly be normalised) that slightly depend on the inflaton potential. 

\tableLFITWOPLIAverage

When a logarithmically flat prior is used for $\sigma_\uend$, the reheating scenarios that are mostly populated, at the prior level, are $1$, $4$ and $7$. Let us recall that, up to different reheating parameters priors, case $1$ corresponds to the single-field limit of the models under consideration. Cases $4$ and $7$, on the other hand, are usually associated with larger $\nS$, smaller $r$ and larger $f_{{}_\mathrm{NL}}$. These are the scenarios that mostly contribute to the averaged Bayesian evidence.
In the third column of tables~\ref{table:HINI:Average} and~\ref{table:LFI2PLI:Average} is given the Bayesian evidence corresponding to situations where $\Gamma_\phi<\Gamma_\sigma<m_\sigma$ (cases $1$, $2$ and $3$), $\Gamma_\sigma<\Gamma_\phi<m_\sigma$ (cases $4$, $5$ and $6$) and $\Gamma_\sigma<m_\sigma<\Gamma_\phi$ (cases $7$, $8$, $9$ and $10$). The last column of these tables contain the global, consolidated Bayesian evidence of the models under consideration. For Higgs inflation, introducing an extra light scalar field decreases the logarithm of the Bayesian evidence of the model by less than $\sim 0.5$, at a level that is ``inconclusive'' according to the Jeffreys scale. This is why we conclude that the best single-field models, of the Plateau type, are generally robust under the introduction of such light scalar degrees of freedom. For models that predict a value for $\nS$ that is too low such as Natural Inflation, one can see that scenarios with an extra light scalar field remain, on average, in the ``inconclusive'' zone of their single-field counterpart and are even very slightly disfavoured. The same holds for large-field models predicting a value for $r$ which is too large such as when the potential is quadratic. When the potential is quartic however, models containing an extra light scalar field are moderately favoured compared to their single-field version, and become only moderately disfavoured compared to the best single-field models such as Higgs Inflation. For Power-Law inflation, the situation is rather unchanged, and all versions of the model remain strongly disfavoured.

\subsection[Prior on $\sigma_\uend$]{Prior on ${\bm \sigma_{\mathbf{end}}}$}
\label{sec:sigmaendprior}
\tableHINIAverageGaussian
The results of the Bayesian analysis when the Gaussian prior~(\ref{eq:sigmaend:GaussianPrior}) is used for $\sigma_\uend$ are given in appendix~\ref{sec:Gaussian}.  
The averaged Bayesian evidence is also displayed in tables~\ref{table:HINI:AverageGaussian} and \ref{table:LFI2PLI:AverageGaussian}, and in appendix~\ref{sec:largefield:average:Gaussian} for other large-field models. At the prior level, one can see that the most populated scenarios are $1$, $2$, $3$ and $6$. As already mentioned, scenarios $1$ and $2$ are close to the single-field limit of the models under consideration. Cases $3$ and $6$, on the other hand, correspond to situations where a second phase of inflation takes place. It is therefore interesting to notice that scenarios corresponding to the original ``curvaton'' setup~\cite{Linde:1996gt,Enqvist:2001zp,Lyth:2001nq,Moroi:2001ct} (scenarios 5 and 8) only represent a few percents of the prior space and may not seem very natural, if a Gaussian prior is adopted for $\sigma_\uend$ (see however the caveats about such a prior choice mentioned in section~\ref{sec:priors}). This also emphasises the importance of not restricting the present work to pure curvaton scenarios but investigating all possibilities. Since situations with extra phases of inflation probe steeper parts of the inflaton potential, they are always disfavoured. This is why we find that the averaged Bayesian evidence of all potentials considered in this section is decreased when an extra light scalar field is present, and that the corresponding models are ``weakly disfavoured'' compared to their single-field counterpart.

\tableLFITWOPLIAverageGaussian

\section{Conclusion}
\label{sec:conclusion}

Let us now summarise our main results. In this paper, we have used Bayesian inference techniques to investigate situations where an extra light scalar field is present during inflation and reheating. Combining the analytical work of \Ref{Vennin:2015vfa} with the numerical tools developed in \Refs{Ringeval:2013lea,Martin:2013nzq}, we have designed a numerical pipeline  where $\sim 200$ inflaton setups $\times\, 10$ reheating scenarios $= 2000$ models are implemented. For simplicity, we have presented the results obtained for a few prototypical inflaton potentials only, but they are representative of the generic trends that can be more generally observed.

We have found that plateau models, that already provide a good fit to the data in their single-field version, are very robust under the introduction of a light scalar field, which only decreases the Bayesian evidence at an inconclusive level. For hilltop potentials, single-field scenarios usually lead to values of $\nS$ that are too low when the width of the hill is not super-Planckian. When a light scalar field is added, the right value of $\nS$ can be obtained, but this happens for very fine-tuned values of the extra field parameters and/or when large non-Gaussianities are produced. As a consequence, the Bayesian status of these models is not improved. The same holds for most large-field models, that give rise to values of $r$ that are too large in their single-field version. The only exception is quartic potentials, for which we have found that reheating scenarios $5$ and $8$ are ``strongly favoured'' with respect to their single-field counterpart, and that scenario $5$ even lies in the inconclusive zone of (the single-field version of) Higgs Inflation (the Starobinsky model). 

To summarise, we have thus showed that the best models after Planck are of two kinds: plateau potentials, regardless of whether an extra field is added or not, and quartic large-field inflation with an extra light scalar field, in reheating scenarios $5$ and $8$.

The role non-Gaussianities play in constraining these models~\cite{Valiviita:2006mz, Huang:2008ze, Enqvist:2008gk, Nakayama:2009ce, Byrnes:2010xd, Enomoto:2012uy, Fonseca:2012cj, Hardwick:2015tma} was also discussed. We have seen that current bounds on non-Gaussianities are already sufficient to constrain hilltop and large-field potentials, but not plateau potentials. In the future, an important question is therefore whether the most efficient way to constrain these models is to improve measurements on $\nS$ and $r$ or measurements on non-Gaussianities. We plan to address this issue in a separate paper.

Different priors have also been compared for the \textit{vev} of the added field at the end of inflation. When it is set to its expected value from quantum dispersion effects, we have found that the most natural scenarios are the ones containing extra phases of inflation, that are generally disfavoured. This also means that the results of our analysis are sensitive to the distribution chosen for the \textit{vev} of the light field, and that given a model, such a distribution can therefore be constrained. This may be relevant to the question~\cite{Starobinsky:1986fx, Enqvist:2012, Burgess:2015ajz} whether observations can give access to scales beyond the observational horizon.

Finally, we have used Bayesian complexity to quantify the number of parameters that are left unconstrained by the present analysis. We have found that, while the models studied here have more unconstrained parameters than their single-field versions, they also allow us to constrain more parameters. This is due to the fact that the added parameters not only contribute to the kinematic description of reheating but also to the statistical properties of the perturbations themselves, to which the added field contributes. The interplay between these two effects lead to a suppression of degeneracies that is responsible for having more constrained parameters. This also means that non trivial constraints on the reheating temperatures can be obtained in these models, which we plan to investigate in a future publication.

\section*{Acknowledgments}
It is a pleasure to thank Christophe Ringeval for sharing his effective likelihood code with us and Robert Hardwick for careful reading of this manuscript. This work was realised using the ICG Sciama HPC cluster and we thank Gary Burton for constant help. This work is supported by STFC grants ST/K00090X/1 and ST/L005573/1.
\clearpage
\appendix
\section{Reheating Cases}
\label{sec:ReheatingCases}
\begin{figure*}[h]
\begin{center}
\includegraphics[width=0.99\textwidth]{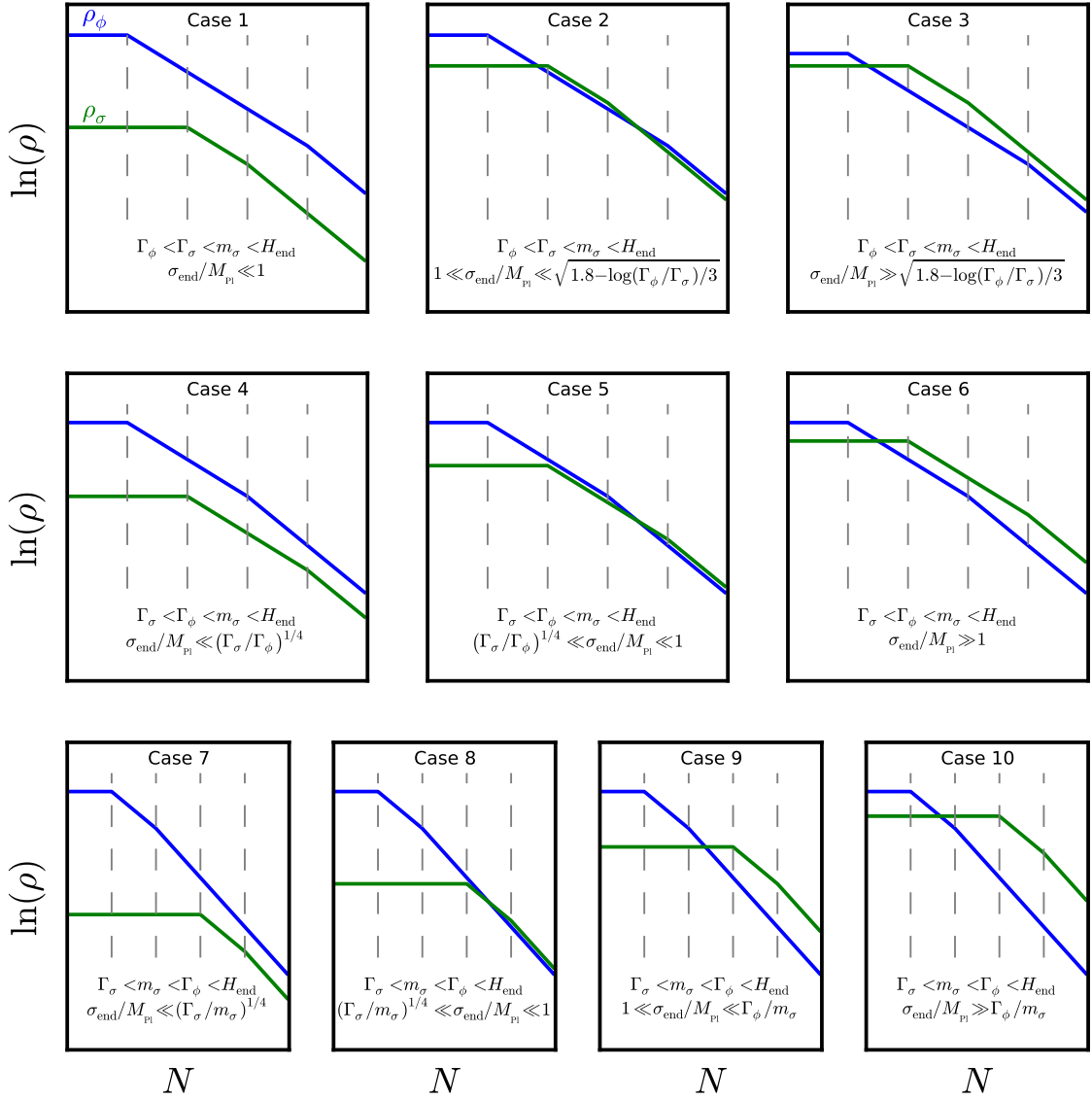}
\caption{Different possible reheating scenarios, depending on the values taken by $\Gamma_\sigma$, $m_\sigma$, $\Gamma_\phi$, $H_\uend$ and $\sigma_\uend$. Cases 1, 2 and 3 correspond to $\Gamma_\phi<\Gamma_\sigma<m_\sigma<H_\uend$; cases 4, 5 and 6 correspond to $\Gamma_\sigma<\Gamma_\phi<m_\sigma<H_\uend$; cases 7, 8, 9 and 10 correspond to $\Gamma_\sigma<m_\sigma<\Gamma_\phi<H_\uend$. Within each row, different cases are distinguished by $\sigma_\uend/\Mp$ which controls when $\sigma$ dominates the total energy density (the precise values for $\sigma_\uend$ at the limit between the different scenarios are given in \Ref{Vennin:2015vfa}). The blue curves stand for the energy density of $\phi$ while the green ones are for $\sigma$.}
\label{fig:rho}
\end{center}
\end{figure*}
\newpage
\section{Bayesian Results for Other Large-Field Models}
\label{sec:lfi:otherp}
In the following tables, we display the results of the Bayesian analysis when the inflaton potential is given by \Eq{eq:lfi:pot} with $p=2/3$, $p=1$, $p=2$, $p=3$ and $p=4$, as well as for marginalising over $p\in [0.2,5]$.
\begin{multicols}{2}
\noindent
\begin{baytabular}
$\lfi$ & $\Elfi$ & $\NPlfi$ & $\NUPlfi$ & $\BElfi$ \\
\hline\hline
$\mclfiONE$ & $\EmclfiONE$ & $\NPmclfiONE$ & $\NUPmclfiONE$ & $\BEmclfiONE$ \\
\hline
$\mclfiTWO$ & $\EmclfiTWO$ & $\NPmclfiTWO$ & $\NUPmclfiTWO$ & $\BEmclfiTWO$ \\
\hline
$\mclfiTHREE$ & $\EmclfiTHREE$ & $\NPmclfiTHREE$ & $\NUPmclfiTHREE$ & $\BEmclfiTHREE$ \\
\hline
$\mclfiFOUR$ & $\EmclfiFOUR$ & $\NPmclfiFOUR$ & $\NUPmclfiFOUR$ & $\BEmclfiFOUR$ \\
\hline
$\mclfiFIVE$ & $\EmclfiFIVE$ & $\NPmclfiFIVE$ & $\NUPmclfiFIVE$ & $\BEmclfiFIVE$ \\
\hline
$\mclfiSIX$ & $\EmclfiSIX$ & $\NPmclfiSIX$ & $\NUPmclfiSIX$ & $\BEmclfiSIX$ \\
\hline
$\mclfiSEVEN$ & $\EmclfiSEVEN$ & $\NPmclfiSEVEN$ & $\NUPmclfiSEVEN$ & $\BEmclfiSEVEN$ \\
\hline
$\mclfiEIGHT$ & $\EmclfiEIGHT$ & $\NPmclfiEIGHT$ & $\NUPmclfiEIGHT$ & $\BEmclfiEIGHT$ \\
\hline
$\mclfiNINE$ & $\EmclfiNINE$ & $\NPmclfiNINE$ & $\NUPmclfiNINE$ & $\BEmclfiNINE$ \\
\hline
$\mclfiONEZERO$ & $\EmclfiONEZERO$ & $\NPmclfiONEZERO$ & $\NUPmclfiONEZERO$ & $\BEmclfiONEZERO$ \\
\hline
\end{baytabular}
\noindent
\begin{baytabular}
$\lfiTWOTHREE$ & $\ElfiTWOTHREE$ & $\NPlfiTWOTHREE$ & $\NUPlfiTWOTHREE$ & $\BElfiTWOTHREE$ \\
\hline\hline
$\mclfiONETWOTHREE$ & $\EmclfiONETWOTHREE$ & $\NPmclfiONETWOTHREE$ & $\NUPmclfiONETWOTHREE$ & $\BEmclfiONETWOTHREE$ \\
\hline
$\mclfiTWOTWOTHREE$ & $\EmclfiTWOTWOTHREE$ & $\NPmclfiTWOTWOTHREE$ & $\NUPmclfiTWOTWOTHREE$ & $\BEmclfiTWOTWOTHREE$ \\
\hline
$\mclfiTHREETWOTHREE$ & $\EmclfiTHREETWOTHREE$ & $\NPmclfiTHREETWOTHREE$ & $\NUPmclfiTHREETWOTHREE$ & $\BEmclfiTHREETWOTHREE$ \\
\hline
$\mclfiFOURTWOTHREE$ & $\EmclfiFOURTWOTHREE$ & $\NPmclfiFOURTWOTHREE$ & $\NUPmclfiFOURTWOTHREE$ & $\BEmclfiFOURTWOTHREE$ \\
\hline
$\mclfiFIVETWOTHREE$ & $\EmclfiFIVETWOTHREE$ & $\NPmclfiFIVETWOTHREE$ & $\NUPmclfiFIVETWOTHREE$ & $\BEmclfiFIVETWOTHREE$ \\
\hline
$\mclfiSIXTWOTHREE$ & $\EmclfiSIXTWOTHREE$ & $\NPmclfiSIXTWOTHREE$ & $\NUPmclfiSIXTWOTHREE$ & $\BEmclfiSIXTWOTHREE$ \\
\hline
$\mclfiSEVENTWOTHREE$ & $\EmclfiSEVENTWOTHREE$ & $\NPmclfiSEVENTWOTHREE$ & $\NUPmclfiSEVENTWOTHREE$ & $\BEmclfiSEVENTWOTHREE$ \\
\hline
$\mclfiEIGHTTWOTHREE$ & $\EmclfiEIGHTTWOTHREE$ & $\NPmclfiEIGHTTWOTHREE$ & $\NUPmclfiEIGHTTWOTHREE$ & $\BEmclfiEIGHTTWOTHREE$ \\
\hline
$\mclfiNINETWOTHREE$ & $\EmclfiNINETWOTHREE$ & $\NPmclfiNINETWOTHREE$ & $\NUPmclfiNINETWOTHREE$ & $\BEmclfiNINETWOTHREE$ \\
\hline
$\mclfiONEZEROTWOTHREE$ & $\EmclfiONEZEROTWOTHREE$ & $\NPmclfiONEZEROTWOTHREE$ & $\NUPmclfiONEZEROTWOTHREE$ & $\BEmclfiONEZEROTWOTHREE$ \\
\hline
\end{baytabular}
\noindent 
\end{multicols}

\begin{multicols}{2}
\noindent
\begin{baytabular}
$\lfiONE$ & $\ElfiONE$ & $\NPlfiONE$ & $\NUPlfiONE$ & $\BElfiONE$ \\
\hline\hline
$\mclfiONEONE$ & $\EmclfiONEONE$ & $\NPmclfiONEONE$ & $\NUPmclfiONEONE$ & $\BEmclfiONEONE$ \\
\hline
$\mclfiTWOONE$ & $\EmclfiTWOONE$ & $\NPmclfiTWOONE$ & $\NUPmclfiTWOONE$ & $\BEmclfiTWOONE$ \\
\hline
$\mclfiTHREEONE$ & $\EmclfiTHREEONE$ & $\NPmclfiTHREEONE$ & $\NUPmclfiTHREEONE$ & $\BEmclfiTHREEONE$ \\
\hline
$\mclfiFOURONE$ & $\EmclfiFOURONE$ & $\NPmclfiFOURONE$ & $\NUPmclfiFOURONE$ & $\BEmclfiFOURONE$ \\
\hline
$\mclfiFIVEONE$ & $\EmclfiFIVEONE$ & $\NPmclfiFIVEONE$ & $\NUPmclfiFIVEONE$ & $\BEmclfiFIVEONE$ \\
\hline
$\mclfiSIXONE$ & $\EmclfiSIXONE$ & $\NPmclfiSIXONE$ & $\NUPmclfiSIXONE$ & $\BEmclfiSIXONE$ \\
\hline
$\mclfiSEVENONE$ & $\EmclfiSEVENONE$ & $\NPmclfiSEVENONE$ & $\NUPmclfiSEVENONE$ & $\BEmclfiSEVENONE$ \\
\hline
$\mclfiEIGHTONE$ & $\EmclfiEIGHTONE$ & $\NPmclfiEIGHTONE$ & $\NUPmclfiEIGHTONE$ & $\BEmclfiEIGHTONE$ \\
\hline
$\mclfiNINEONE$ & $\EmclfiNINEONE$ & $\NPmclfiNINEONE$ & $\NUPmclfiNINEONE$ & $\BEmclfiNINEONE$ \\
\hline
$\mclfiONEZEROONE$ & $\EmclfiONEZEROONE$ & $\NPmclfiONEZEROONE$ & $\NUPmclfiONEZEROONE$ & $\BEmclfiONEZEROONE$ \\
\hline
\end{baytabular}
\noindent
\begin{baytabular}
$\lfiTWO$ & $\ElfiTWO$ & $\NPlfiTWO$ & $\NUPlfiTWO$ & $\BElfiTWO$ \\
\hline\hline
$\mclfiONETWO$ & $\EmclfiONETWO$ & $\NPmclfiONETWO$ & $\NUPmclfiONETWO$ & $\BEmclfiONETWO$ \\
\hline
$\mclfiTWOTWO$ & $\EmclfiTWOTWO$ & $\NPmclfiTWOTWO$ & $\NUPmclfiTWOTWO$ & $\BEmclfiTWOTWO$ \\
\hline
$\mclfiTHREETWO$ & $\EmclfiTHREETWO$ & $\NPmclfiTHREETWO$ & $\NUPmclfiTHREETWO$ & $\BEmclfiTHREETWO$ \\
\hline
$\mclfiFOURTWO$ & $\EmclfiFOURTWO$ & $\NPmclfiFOURTWO$ & $\NUPmclfiFOURTWO$ & $\BEmclfiFOURTWO$ \\
\hline
$\mclfiFIVETWO$ & $\EmclfiFIVETWO$ & $\NPmclfiFIVETWO$ & $\NUPmclfiFIVETWO$ & $\BEmclfiFIVETWO$ \\
\hline
$\mclfiSIXTWO$ & $\EmclfiSIXTWO$ & $\NPmclfiSIXTWO$ & $\NUPmclfiSIXTWO$ & $\BEmclfiSIXTWO$ \\
\hline
$\mclfiSEVENTWO$ & $\EmclfiSEVENTWO$ & $\NPmclfiSEVENTWO$ & $\NUPmclfiSEVENTWO$ & $\BEmclfiSEVENTWO$ \\
\hline
$\mclfiEIGHTTWO$ & $\EmclfiEIGHTTWO$ & $\NPmclfiEIGHTTWO$ & $\NUPmclfiEIGHTTWO$ & $\BEmclfiEIGHTTWO$ \\
\hline
$\mclfiNINETWO$ & $\EmclfiNINETWO$ & $\NPmclfiNINETWO$ & $\NUPmclfiNINETWO$ & $\BEmclfiNINETWO$ \\
\hline
$\mclfiONEZEROTWO$ & $\EmclfiONEZEROTWO$ & $\NPmclfiONEZEROTWO$ & $\NUPmclfiONEZEROTWO$ & $\BEmclfiONEZEROTWO$ \\
\hline
\end{baytabular}
\end{multicols}

\begin{multicols}{2}
\noindent
\begin{baytabular}
$\lfiTHREE$ & $\ElfiTHREE$ & $\NPlfiTHREE$ & $\NUPlfiTHREE$ & $\BElfiTHREE$ \\
\hline\hline
$\mclfiONETHREE$ & $\EmclfiONETHREE$ & $\NPmclfiONETHREE$ & $\NUPmclfiONETHREE$ & $\BEmclfiONETHREE$ \\
\hline
$\mclfiTWOTHREE$ & $\EmclfiTWOTHREE$ & $\NPmclfiTWOTHREE$ & $\NUPmclfiTWOTHREE$ & $\BEmclfiTWOTHREE$ \\
\hline
$\mclfiTHREETHREE$ & $\EmclfiTHREETHREE$ & $\NPmclfiTHREETHREE$ & $\NUPmclfiTHREETHREE$ & $\BEmclfiTHREETHREE$ \\
\hline
$\mclfiFOURTHREE$ & $\EmclfiFOURTHREE$ & $\NPmclfiFOURTHREE$ & $\NUPmclfiFOURTHREE$ & $\BEmclfiFOURTHREE$ \\
\hline
$\mclfiFIVETHREE$ & $\EmclfiFIVETHREE$ & $\NPmclfiFIVETHREE$ & $\NUPmclfiFIVETHREE$ & $\BEmclfiFIVETHREE$ \\
\hline
$\mclfiSIXTHREE$ & $\EmclfiSIXTHREE$ & $\NPmclfiSIXTHREE$ & $\NUPmclfiSIXTHREE$ & $\BEmclfiSIXTHREE$ \\
\hline
$\mclfiSEVENTHREE$ & $\EmclfiSEVENTHREE$ & $\NPmclfiSEVENTHREE$ & $\NUPmclfiSEVENTHREE$ & $\BEmclfiSEVENTHREE$ \\
\hline
$\mclfiEIGHTTHREE$ & $\EmclfiEIGHTTHREE$ & $\NPmclfiEIGHTTHREE$ & $\NUPmclfiEIGHTTHREE$ & $\BEmclfiEIGHTTHREE$ \\
\hline
$\mclfiNINETHREE$ & $\EmclfiNINETHREE$ & $\NPmclfiNINETHREE$ & $\NUPmclfiNINETHREE$ & $\BEmclfiNINETHREE$ \\
\hline
$\mclfiONEZEROTHREE$ & $\EmclfiONEZEROTHREE$ & $\NPmclfiONEZEROTHREE$ & $\NUPmclfiONEZEROTHREE$ & $\BEmclfiONEZEROTHREE$ \\
\hline
\end{baytabular}
\noindent
\begin{baytabular}
$\lfiFOUR$ & $\ElfiFOUR$ & $\NPlfiFOUR$ & $\NUPlfiFOUR$ & $\BElfiFOUR$ \\
\hline\hline
$\mclfiONEFOUR$ & $\EmclfiONEFOUR$ & $\NPmclfiONEFOUR$ & $\NUPmclfiONEFOUR$ & $\BEmclfiONEFOUR$ \\
\hline
$\mclfiTWOFOUR$ & $\EmclfiTWOFOUR$ & $\NPmclfiTWOFOUR$ & $\NUPmclfiTWOFOUR$ & $\BEmclfiTWOFOUR$ \\
\hline
$\mclfiTHREEFOUR$ & $\EmclfiTHREEFOUR$ & $\NPmclfiTHREEFOUR$ & $\NUPmclfiTHREEFOUR$ & $\BEmclfiTHREEFOUR$ \\
\hline
$\mclfiFOURFOUR$ & $\EmclfiFOURFOUR$ & $\NPmclfiFOURFOUR$ & $\NUPmclfiFOURFOUR$ & $\BEmclfiFOURFOUR$ \\
\hline
$\mclfiFIVEFOUR$ & $\EmclfiFIVEFOUR$ & $\NPmclfiFIVEFOUR$ & $\NUPmclfiFIVEFOUR$ & $\BEmclfiFIVEFOUR$ \\
\hline
$\mclfiSIXFOUR$ & $\EmclfiSIXFOUR$ & $\NPmclfiSIXFOUR$ & $\NUPmclfiSIXFOUR$ & $\BEmclfiSIXFOUR$ \\
\hline
$\mclfiSEVENFOUR$ & $\EmclfiSEVENFOUR$ & $\NPmclfiSEVENFOUR$ & $\NUPmclfiSEVENFOUR$ & $\BEmclfiSEVENFOUR$ \\
\hline
$\mclfiEIGHTFOUR$ & $\EmclfiEIGHTFOUR$ & $\NPmclfiEIGHTFOUR$ & $\NUPmclfiEIGHTFOUR$ & $\BEmclfiEIGHTFOUR$ \\
\hline
$\mclfiNINEFOUR$ & $\EmclfiNINEFOUR$ & $\NPmclfiNINEFOUR$ & $\NUPmclfiNINEFOUR$ & $\BEmclfiNINEFOUR$ \\
\hline
$\mclfiONEZEROFOUR$ & $\EmclfiONEZEROFOUR$ & $\NPmclfiONEZEROFOUR$ & $\NUPmclfiONEZEROFOUR$ & $\BEmclfiONEZEROFOUR$ \\
\hline
\end{baytabular}
\end{multicols}
\newpage
\section{Averaging over Reheating Scenarios for Other Large-Field Models}
\label{sec:largefield:average}

\tableLFIAverageONE
\tableLFIAverageTWO
\tableLFIAverageTHREE

\newpage
\section{Bayesian Results with a Gaussian Prior on $\sigma_\uend$}
\label{sec:Gaussian}
In this appendix, we give the Bayesian evidence and number of unconstrained parameters in the case where a ``Gaussian'' (in the sense defined in section~\ref{sec:priors}) prior for $\sigma_\uend$ is chosen. The value of the maximum likelihood $\mathcal{L}_\umax$ is not reported since it is independent of the prior, and is therefore the same as the one given in the tables of section~\ref{sec:results} and appendix~\ref{sec:lfi:otherp}.

\begin{multicols}{2}
\begin{center}
\noindent
\begin{baytabularGaussian}
$\hi$ & $\EhiGaussian$ & $\NPhi$ & $\NUPhiGaussian$  \\
\hline\hline
$\mchiONE$ & $\EmchiONEGaussian$ & $\NPmchiONE$ & $\NUPmchiONEGaussian$  \\
\hline
$\mchiTWO$ & $\EmchiTWOGaussian$ & $\NPmchiTWO$ & $\NUPmchiTWOGaussian$  \\
\hline
$\mchiTHREE$ & $\EmchiTHREEGaussian$ & $\NPmchiTHREE$ & $\NUPmchiTHREEGaussian$  \\
\hline
$\mchiFOUR$ & $\EmchiFOURGaussian$ & $\NPmchiFOUR$ & $\NUPmchiFOURGaussian$  \\
\hline
$\mchiFIVE$ & $\EmchiFIVEGaussian$ & $\NPmchiFIVE$ & $\NUPmchiFIVEGaussian$  \\
\hline
$\mchiSIX$ & $\EmchiSIXGaussian$ & $\NPmchiSIX$ & $\NUPmchiSIXGaussian$  \\
\hline
$\mchiSEVEN$ & $\EmchiSEVENGaussian$ & $\NPmchiSEVEN$ & $\NUPmchiSEVENGaussian$ \\
\hline
$\mchiEIGHT$ & $\EmchiEIGHTGaussian$ & $\NPmchiEIGHT$ & $\NUPmchiEIGHTGaussian$  \\
\hline
$\mchiNINE$ & $\EmchiNINEGaussian$ & $\NPmchiNINE$ & $\NUPmchiNINEGaussian$  \\
\hline
$\mchiONEZERO$ & $\EmchiONEZEROGaussian$ & $\NPmchiONEZERO$ & $\NUPmchiONEZEROGaussian$  \\
\hline
\end{baytabularGaussian}
\noindent
\begin{baytabularGaussian}
$\nati$ & $\EniGaussian$ & $\NPni$ & $\NUPniGaussian$  \\
\hline\hline
$\mcniONE$ & $\EmcniONEGaussian$ & $\NPmcniONE$ & $\NUPmcniONEGaussian$  \\
\hline
$\mcniTWO$ & $\EmcniTWOGaussian$ & $\NPmcniTWO$ & $\NUPmcniTWOGaussian$  \\
\hline
$\mcniTHREE$ & $\EmcniTHREEGaussian$ & $\NPmcniTHREE$ & $\NUPmcniTHREEGaussian$  \\
\hline
$\mcniFOUR$ & $\EmcniFOURGaussian$ & $\NPmcniFOUR$ & $\NUPmcniFOURGaussian$  \\
\hline
$\mcniFIVE$ & $\EmcniFIVEGaussian$ & $\NPmcniFIVE$ & $\NUPmcniFIVEGaussian$  \\
\hline
$\mcniSIX$ & $\EmcniSIXGaussian$ & $\NPmcniSIX$ & $\NUPmcniSIXGaussian$  \\
\hline
$\mcniSEVEN$ & $\EmcniSEVENGaussian$ & $\NPmcniSEVEN$ & $\NUPmcniSEVENGaussian$  \\
\hline
$\mcniEIGHT$ & $\EmcniEIGHTGaussian$ & $\NPmcniEIGHT$ & $\NUPmcniEIGHTGaussian$  \\
\hline
$\mcniNINE$ & $\EmcniNINEGaussian$ & $\NPmcniNINE$ & $\NUPmcniNINEGaussian$  \\
\hline
$\mcniONEZERO$ & $\EmcniONEZEROGaussian$ & $\NPmcniONEZERO$ & $\NUPmcniONEZEROGaussian$  \\
\hline
\end{baytabularGaussian}
\end{center}
\end{multicols}

\begin{multicols}{2}
\begin{center}
\noindent
\begin{baytabularGaussian}
$\lfiTWO$ & $\ElfiTWOGaussian$ & $\NPlfiTWO$ & $\NUPlfiTWOGaussian$  \\
\hline\hline
$\mclfiONETWO$ & $\EmclfiONETWOGaussian$ & $\NPmclfiONETWO$ & $\NUPmclfiONETWOGaussian$  \\
\hline
$\mclfiTWOTWO$ & $\EmclfiTWOTWOGaussian$ & $\NPmclfiTWOTWO$ & $\NUPmclfiTWOTWOGaussian$  \\
\hline
$\mclfiTHREETWO$ & $\EmclfiTHREETWOGaussian$ & $\NPmclfiTHREETWO$ & $\NUPmclfiTHREETWOGaussian$  \\
\hline
$\mclfiFOURTWO$ & $\EmclfiFOURTWOGaussian$ & $\NPmclfiFOURTWO$ & $\NUPmclfiFOURTWOGaussian$  \\
\hline
$\mclfiFIVETWO$ & $\EmclfiFIVETWOGaussian$ & $\NPmclfiFIVETWO$ & $\NUPmclfiFIVETWOGaussian$  \\
\hline
$\mclfiSIXTWO$ & $\EmclfiSIXTWOGaussian$ & $\NPmclfiSIXTWO$ & $\NUPmclfiSIXTWOGaussian$  \\
\hline
$\mclfiSEVENTWO$ & $\EmclfiSEVENTWOGaussian$ & $\NPmclfiSEVENTWO$ & $\NUPmclfiSEVENTWOGaussian$  \\
\hline
$\mclfiEIGHTTWO$ & $\EmclfiEIGHTTWOGaussian$ & $\NPmclfiEIGHTTWO$ & $\NUPmclfiEIGHTTWOGaussian$  \\
\hline
$\mclfiNINETWO$ & $\EmclfiNINETWOGaussian$ & $\NPmclfiNINETWO$ & $\NUPmclfiNINETWOGaussian$  \\
\hline
$\mclfiONEZEROTWO$ & $\EmclfiONEZEROTWOGaussian$ & $\NPmclfiONEZEROTWO$ & $\NUPmclfiONEZEROTWOGaussian$  \\
\hline
\end{baytabularGaussian}
\noindent
\begin{baytabularGaussian}
$\pli$ & $\EpliGaussian$ & $\NPpli$ & $\NUPpliGaussian$  \\
\hline\hline
$\mcpliONE$ & $\EmcpliONEGaussian$ & $\NPmcpliONE$ & $\NUPmcpliONEGaussian$  \\
\hline
$\mcpliTWO$ & $\EmcpliTWOGaussian$ & $\NPmcpliTWO$ & $\NUPmcpliTWOGaussian$  \\
\hline
$\mcpliTHREE$ & $\EmcpliTHREEGaussian$ & $\NPmcpliTHREE$ & $\NUPmcpliTHREEGaussian$ \\
\hline
$\mcpliFOUR$ & $\EmcpliFOURGaussian$ & $\NPmcpliFOUR$ & $\NUPmcpliFOURGaussian$  \\
\hline
$\mcpliFIVE$ & $\EmcpliFIVEGaussian$ & $\NPmcpliFIVE$ & $\NUPmcpliFIVEGaussian$  \\
\hline
$\mcpliSIX$ & $\EmcpliSIXGaussian$ & $\NPmcpliSIX$ & $\NUPmcpliSIXGaussian$  \\
\hline
$\mcpliSEVEN$ & $\EmcpliSEVENGaussian$ & $\NPmcpliSEVEN$ & $\NUPmcpliSEVENGaussian$  \\
\hline
$\mcpliEIGHT$ & $\EmcpliEIGHTGaussian$ & $\NPmcpliEIGHT$ & $\NUPmcpliEIGHTGaussian$ \\
\hline
$\mcpliNINE$ & $\EmcpliNINEGaussian$ & $\NPmcpliNINE$ & $\NUPmcpliNINEGaussian$  \\
\hline
$\mcpliONEZERO$ & $\EmcpliONEZEROGaussian$ & $\NPmcpliONEZERO$ & $\NUPmcpliONEZEROGaussian$ \\
\hline
\end{baytabularGaussian}
\end{center}
\end{multicols}

\begin{multicols}{2}
\begin{center}
\noindent
\begin{baytabularGaussian}
$\lfi$ & $\ElfiGaussian$ & $\NPlfi$ & $\NUPlfiGaussian$   \\
\hline\hline
$\mclfiONE$ & $\EmclfiONEGaussian$ & $\NPmclfiONE$ & $\NUPmclfiONEGaussian$   \\
\hline
$\mclfiTWO$ & $\EmclfiTWOGaussian$ & $\NPmclfiTWO$ & $\NUPmclfiTWOGaussian$   \\
\hline
$\mclfiTHREE$ & $\EmclfiTHREEGaussian$ & $\NPmclfiTHREE$ & $\NUPmclfiTHREEGaussian$   \\
\hline
$\mclfiFOUR$ & $\EmclfiFOURGaussian$ & $\NPmclfiFOUR$ & $\NUPmclfiFOURGaussian$   \\
\hline
$\mclfiFIVE$ & $\EmclfiFIVEGaussian$ & $\NPmclfiFIVE$ & $\NUPmclfiFIVEGaussian$ \\
\hline
$\mclfiSIX$ & $\EmclfiSIXGaussian$ & $\NPmclfiSIX$ & $\NUPmclfiSIXGaussian$   \\
\hline
$\mclfiSEVEN$ & $\EmclfiSEVENGaussian$ & $\NPmclfiSEVEN$ & $\NUPmclfiSEVENGaussian$   \\
\hline
$\mclfiEIGHT$ & $\EmclfiEIGHTGaussian$ & $\NPmclfiEIGHT$ & $\NUPmclfiEIGHTGaussian$   \\
\hline
$\mclfiNINE$ & $\EmclfiNINEGaussian$ & $\NPmclfiNINE$ & $\NUPmclfiNINEGaussian$   \\
\hline
$\mclfiONEZERO$ & $\EmclfiONEZEROGaussian$ & $\NPmclfiONEZERO$ & $\NUPmclfiONEZEROGaussian$   \\
\hline
\end{baytabularGaussian}
\noindent
\begin{baytabularGaussian}
$\lfiTWOTHREE$ & $\ElfiTWOTHREEGaussian$ & $\NPlfiTWOTHREE$ & $\NUPlfiTWOTHREEGaussian$   \\
\hline\hline
$\mclfiONETWOTHREE$ & $\EmclfiONETWOTHREEGaussian$ & $\NPmclfiONETWOTHREE$ & $\NUPmclfiONETWOTHREEGaussian$   \\
\hline
$\mclfiTWOTWOTHREE$ & $\EmclfiTWOTWOTHREEGaussian$ & $\NPmclfiTWOTWOTHREE$ & $\NUPmclfiTWOTWOTHREEGaussian$   \\
\hline
$\mclfiTHREETWOTHREE$ & $\EmclfiTHREETWOTHREEGaussian$ & $\NPmclfiTHREETWOTHREE$ & $\NUPmclfiTHREETWOTHREEGaussian$  \\
\hline
$\mclfiFOURTWOTHREE$ & $\EmclfiFOURTWOTHREEGaussian$ & $\NPmclfiFOURTWOTHREE$ & $\NUPmclfiFOURTWOTHREEGaussian$   \\
\hline
$\mclfiFIVETWOTHREE$ & $\EmclfiFIVETWOTHREEGaussian$ & $\NPmclfiFIVETWOTHREE$ & $\NUPmclfiFIVETWOTHREEGaussian$   \\
\hline
$\mclfiSIXTWOTHREE$ & $\EmclfiSIXTWOTHREEGaussian$ & $\NPmclfiSIXTWOTHREE$ & $\NUPmclfiSIXTWOTHREEGaussian$   \\
\hline
$\mclfiSEVENTWOTHREE$ & $\EmclfiSEVENTWOTHREEGaussian$ & $\NPmclfiSEVENTWOTHREE$ & $\NUPmclfiSEVENTWOTHREEGaussian$  \\
\hline
$\mclfiEIGHTTWOTHREE$ & $\EmclfiEIGHTTWOTHREEGaussian$ & $\NPmclfiEIGHTTWOTHREE$ & $\NUPmclfiEIGHTTWOTHREEGaussian$   \\
\hline
$\mclfiNINETWOTHREE$ & $\EmclfiNINETWOTHREEGaussian$ & $\NPmclfiNINETWOTHREE$ & $\NUPmclfiNINETWOTHREEGaussian$   \\
\hline
$\mclfiONEZEROTWOTHREE$ & $\EmclfiONEZEROTWOTHREEGaussian$ & $\NPmclfiONEZEROTWOTHREE$ & $\NUPmclfiONEZEROTWOTHREEGaussian$  \\
\hline
\end{baytabularGaussian}
\noindent 
\end{center}
\end{multicols}

\clearpage

\begin{multicols}{2}
\begin{center}
\noindent
\begin{baytabularGaussian}
$\lfiONE$ & $\ElfiONEGaussian$ & $\NPlfiONE$ & $\NUPlfiONEGaussian$   \\
\hline\hline
$\mclfiONEONE$ & $\EmclfiONEONEGaussian$ & $\NPmclfiONEONE$ & $\NUPmclfiONEONEGaussian$   \\
\hline
$\mclfiTWOONE$ & $\EmclfiTWOONEGaussian$ & $\NPmclfiTWOONE$ & $\NUPmclfiTWOONEGaussian$   \\
\hline
$\mclfiTHREEONE$ & $\EmclfiTHREEONEGaussian$ & $\NPmclfiTHREEONE$ & $\NUPmclfiTHREEONEGaussian$  \\
\hline
$\mclfiFOURONE$ & $\EmclfiFOURONEGaussian$ & $\NPmclfiFOURONE$ & $\NUPmclfiFOURONEGaussian$   \\
\hline
$\mclfiFIVEONE$ & $\EmclfiFIVEONEGaussian$ & $\NPmclfiFIVEONE$ & $\NUPmclfiFIVEONEGaussian$  \\
\hline
$\mclfiSIXONE$ & $\EmclfiSIXONEGaussian$ & $\NPmclfiSIXONE$ & $\NUPmclfiSIXONEGaussian$  \\
\hline
$\mclfiSEVENONE$ & $\EmclfiSEVENONEGaussian$ & $\NPmclfiSEVENONE$ & $\NUPmclfiSEVENONEGaussian$   \\
\hline
$\mclfiEIGHTONE$ & $\EmclfiEIGHTONEGaussian$ & $\NPmclfiEIGHTONE$ & $\NUPmclfiEIGHTONEGaussian$   \\
\hline
$\mclfiNINEONE$ & $\EmclfiNINEONEGaussian$ & $\NPmclfiNINEONE$ & $\NUPmclfiNINEONEGaussian$   \\
\hline
$\mclfiONEZEROONE$ & $\EmclfiONEZEROONEGaussian$ & $\NPmclfiONEZEROONE$ & $\NUPmclfiONEZEROONEGaussian$  \\
\hline
\end{baytabularGaussian}
\noindent
\begin{baytabularGaussian}
$\lfiTHREE$ & $\ElfiTHREEGaussian$ & $\NPlfiTHREE$ & $\NUPlfiTHREEGaussian$   \\
\hline\hline
$\mclfiONETHREE$ & $\EmclfiONETHREEGaussian$ & $\NPmclfiONETHREE$ & $\NUPmclfiONETHREEGaussian$   \\
\hline
$\mclfiTWOTHREE$ & $\EmclfiTWOTHREEGaussian$ & $\NPmclfiTWOTHREE$ & $\NUPmclfiTWOTHREEGaussian$   \\
\hline
$\mclfiTHREETHREE$ & $\EmclfiTHREETHREEGaussian$ & $\NPmclfiTHREETHREE$ & $\NUPmclfiTHREETHREEGaussian$   \\
\hline
$\mclfiFOURTHREE$ & $\EmclfiFOURTHREEGaussian$ & $\NPmclfiFOURTHREE$ & $\NUPmclfiFOURTHREEGaussian$   \\
\hline
$\mclfiFIVETHREE$ & $\EmclfiFIVETHREEGaussian$ & $\NPmclfiFIVETHREE$ & $\NUPmclfiFIVETHREEGaussian$   \\
\hline
$\mclfiSIXTHREE$ & $\EmclfiSIXTHREEGaussian$ & $\NPmclfiSIXTHREE$ & $\NUPmclfiSIXTHREEGaussian$   \\
\hline
$\mclfiSEVENTHREE$ & $\EmclfiSEVENTHREEGaussian$ & $\NPmclfiSEVENTHREE$ & $\NUPmclfiSEVENTHREEGaussian$   \\
\hline
$\mclfiEIGHTTHREE$ & $\EmclfiEIGHTTHREEGaussian$ & $\NPmclfiEIGHTTHREE$ & $\NUPmclfiEIGHTTHREEGaussian$   \\
\hline
$\mclfiNINETHREE$ & $\EmclfiNINETHREEGaussian$ & $\NPmclfiNINETHREE$ & $\NUPmclfiNINETHREEGaussian$   \\
\hline
$\mclfiONEZEROTHREE$ & $\EmclfiONEZEROTHREEGaussian$ & $\NPmclfiONEZEROTHREE$ & $\NUPmclfiONEZEROTHREEGaussian$   \\
\hline
\end{baytabularGaussian}
\end{center}
\end{multicols}

\begin{multicols}{2}
\begin{center}
\noindent
\begin{baytabularGaussian}
$\lfiFOUR$ & $\ElfiFOURGaussian$ & $\NPlfiFOUR$ & $\NUPlfiFOURGaussian$   \\
\hline\hline
$\mclfiONEFOUR$ & $\EmclfiONEFOURGaussian$ & $\NPmclfiONEFOUR$ & $\NUPmclfiONEFOURGaussian$   \\
\hline
$\mclfiTWOFOUR$ & $\EmclfiTWOFOURGaussian$ & $\NPmclfiTWOFOUR$ & $\NUPmclfiTWOFOURGaussian$   \\
\hline
$\mclfiTHREEFOUR$ & $\EmclfiTHREEFOURGaussian$ & $\NPmclfiTHREEFOUR$ & $\NUPmclfiTHREEFOURGaussian$   \\
\hline
$\mclfiFOURFOUR$ & $\EmclfiFOURFOURGaussian$ & $\NPmclfiFOURFOUR$ & $\NUPmclfiFOURFOURGaussian$   \\
\hline
$\mclfiFIVEFOUR$ & $\EmclfiFIVEFOURGaussian$ & $\NPmclfiFIVEFOUR$ & $\NUPmclfiFIVEFOURGaussian$   \\
\hline
$\mclfiSIXFOUR$ & $\EmclfiSIXFOURGaussian$ & $\NPmclfiSIXFOUR$ & $\NUPmclfiSIXFOURGaussian$   \\
\hline
$\mclfiSEVENFOUR$ & $\EmclfiSEVENFOURGaussian$ & $\NPmclfiSEVENFOUR$ & $\NUPmclfiSEVENFOURGaussian$   \\
\hline
$\mclfiEIGHTFOUR$ & $\EmclfiEIGHTFOURGaussian$ & $\NPmclfiEIGHTFOUR$ & $\NUPmclfiEIGHTFOURGaussian$   \\
\hline
$\mclfiNINEFOUR$ & $\EmclfiNINEFOURGaussian$ & $\NPmclfiNINEFOUR$ & $\NUPmclfiNINEFOURGaussian$  \\
\hline
$\mclfiONEZEROFOUR$ & $\EmclfiONEZEROFOURGaussian$ & $\NPmclfiONEZEROFOUR$ & $\NUPmclfiONEZEROFOURGaussian$  \\
\hline
\end{baytabularGaussian}
\noindent
\end{center}
\end{multicols}

\section{Averaging over Reheating Scenarios for Other Large-Field Models with a Gaussian Prior on $\sigma_\uend$}
\label{sec:largefield:average:Gaussian}

\tableLFIAverageONEGaussian
\tableLFIAverageTWOGaussian
\tableLFIAverageTHREEGaussian

\clearpage
\bibliographystyle{JHEP}
\bibliography{curvevid}
\end{document}

%% file: newcommands.tex
\DeclareMathOperator{\erf}{erf}

\newcommand{\ASPIC}{\texttt{ASPIC}\xspace}

\newcommand{\MULTINEST}{\texttt{MultiNest}\xspace}

\newcommand{\dd}{\mathrm{d}}
\newcommand{\ee}{e}

\newcommand{\sss}[1]{{\scriptscriptstyle{#1}}}

\newcommand{\uPl}{\mathrm{Pl}}

\newcommand{\umin}{\mathrm{min}}
\newcommand{\umax}{\mathrm{max}}
\newcommand{\uend}{\mathrm{end}}

\newcommand{\ureh}{\mathrm{reh}}

\newcommand{\uS}{\mathrm{S}}

\newcommand{\usssS}{\sss{\uS}}

\newcommand{\usssPl}{\sss{\uPl}}

\newcommand{\nS}{n_\usssS}

\newcommand{\MeV}{\mathrm{MeV}}

\newcommand{\Mp}{M_\usssPl}

\newcommand{\efolds}{$e$-folds~}

\newcommand{\beq}{\begin{equation}}
\newcommand{\eeq}{\end{equation}}
\newcommand{\bea}{\begin{eqnarray}}
\newcommand{\eea}{\end{eqnarray}}

\newlength{\wsingfig}
\newlength{\wdblefig}
\newlength{\wappfig}
\newlength{\wquadfig}
\newlength{\wtriplefig}
\newcommand{\Eq}[1]{Eq.~(\ref{#1})}

\newcommand{\Fig}[1]{Fig.~{\ref{#1}}}

\newcommand{\Ref}[1]{Ref.~{\cite{#1}}}
\newcommand{\Refs}[1]{Refs.~{\cite{#1}}}

%% file: tables.tex
\newcommand{\tableHINI}{
\begin{table}[t]
\begin{center}
\begin{multicols}{2}
\noindent
\linebreak
\begin{baytabular}
$\hi$ & $\Ehi$ & $\NPhi$ & $\NUPhi$ & $\BEhi$ \\
\hline\hline
$\mchiONE$  & $\EmchiONE$ & $\NPmchiONE$ & $\NUPmchiONE$ & $\BEmchiONE$ \\
\hline
$\mchiTWO$ & $\EmchiTWO$ & $\NPmchiTWO$ & $\NUPmchiTWO$ & $\BEmchiTWO$ \\
\hline
$\mchiTHREE$ & $\EmchiTHREE$ & $\NPmchiTHREE$ & $\NUPmchiTHREE$ & $\BEmchiTHREE$ \\
\hline
$\mchiFOUR$ & $\EmchiFOUR$ & $\NPmchiFOUR$ & $\NUPmchiFOUR$ & $\BEmchiFOUR$ \\
\hline
$\mchiFIVE$ & $\EmchiFIVE$ & $\NPmchiFIVE$ & $\NUPmchiFIVE$ & $\BEmchiFIVE$ \\
\hline
$\mchiSIX$ & $\EmchiSIX$ & $\NPmchiSIX$ & $\NUPmchiSIX$ & $\BEmchiSIX$ \\
\hline
$\mchiSEVEN$ & $\EmchiSEVEN$ & $\NPmchiSEVEN$ & $\NUPmchiSEVEN$ & $\BEmchiSEVEN$ \\
\hline
$\mchiEIGHT$ & $\EmchiEIGHT$ & $\NPmchiEIGHT$ & $\NUPmchiEIGHT$ & $\BEmchiEIGHT$ \\
\hline
$\mchiNINE$ & $\EmchiNINE$ & $\NPmchiNINE$ & $\NUPmchiNINE$ & $\BEmchiNINE$ \\
\hline
$\mchiONEZERO$ & $\EmchiONEZERO$ & $\NPmchiONEZERO$ & $\NUPmchiONEZERO$ & $\BEmchiONEZERO$ \\
\hline
\end{baytabular}
\vfill
\columnbreak
\vspace*{\fill}
\noindent
\begin{baytabular}
$\nati$ & $\Eni$ & $\NPni$ & $\NUPni$ & $\BEni$ \\
\hline\hline
$\mcniONE$ & $\EmcniONE$ & $\NPmcniONE$ & $\NUPmcniONE$ & $\BEmcniONE$ \\
\hline
$\mcniTWO$ & $\EmcniTWO$ & $\NPmcniTWO$ & $\NUPmcniTWO$ & $\BEmcniTWO$ \\
\hline
$\mcniTHREE$ & $\EmcniTHREE$ & $\NPmcniTHREE$ & $\NUPmcniTHREE$ & $\BEmcniTHREE$ \\
\hline
$\mcniFOUR$ & $\EmcniFOUR$ & $\NPmcniFOUR$ & $\NUPmcniFOUR$ & $\BEmcniFOUR$ \\
\hline
$\mcniFIVE$ & $\EmcniFIVE$ & $\NPmcniFIVE$ & $\NUPmcniFIVE$ & $\BEmcniFIVE$ \\
\hline
$\mcniSIX$ & $\EmcniSIX$ & $\NPmcniSIX$ & $\NUPmcniSIX$ & $\BEmcniSIX$ \\
\hline
$\mcniSEVEN$ & $\EmcniSEVEN$ & $\NPmcniSEVEN$ & $\NUPmcniSEVEN$ & $\BEmcniSEVEN$ \\
\hline
$\mcniEIGHT$ & $\EmcniEIGHT$ & $\NPmcniEIGHT$ & $\NUPmcniEIGHT$ & $\BEmcniEIGHT$ \\
\hline
$\mcniNINE$ & $\EmcniNINE$ & $\NPmcniNINE$ & $\NUPmcniNINE$ & $\BEmcniNINE$ \\
\hline
$\mcniONEZERO$ & $\EmcniONEZERO$ & $\NPmcniONEZERO$ & $\NUPmcniONEZERO$ & $\BEmcniONEZERO$ \\
\hline
\end{baytabular}
\end{multicols}
\end{center}
\caption{Bayesian analysis for the Higgs Inflation and Natural Inflation models.}
\label{table:HINI}
\end{table}
}

\newcommand{\tableLFITWOPLI}{
\begin{table}[t]
\begin{center}
\begin{multicols}{2}
\noindent
\begin{baytabular}
$\lfiTWO$ & $\ElfiTWO$ & $\NPlfiTWO$ & $\NUPlfiTWO$ & $\BElfiTWO$ \\
\hline\hline
$\mclfiONETWO$ & $\EmclfiONETWO$ & $\NPmclfiONETWO$ & $\NUPmclfiONETWO$ & $\BEmclfiONETWO$ \\
\hline
$\mclfiTWOTWO$ & $\EmclfiTWOTWO$ & $\NPmclfiTWOTWO$ & $\NUPmclfiTWOTWO$ & $\BEmclfiTWOTWO$ \\
\hline
$\mclfiTHREETWO$ & $\EmclfiTHREETWO$ & $\NPmclfiTHREETWO$ & $\NUPmclfiTHREETWO$ & $\BEmclfiTHREETWO$ \\
\hline
$\mclfiFOURTWO$ & $\EmclfiFOURTWO$ & $\NPmclfiFOURTWO$ & $\NUPmclfiFOURTWO$ & $\BEmclfiFOURTWO$ \\
\hline
$\mclfiFIVETWO$ & $\EmclfiFIVETWO$ & $\NPmclfiFIVETWO$ & $\NUPmclfiFIVETWO$ & $\BEmclfiFIVETWO$ \\
\hline
$\mclfiSIXTWO$ & $\EmclfiSIXTWO$ & $\NPmclfiSIXTWO$ & $\NUPmclfiSIXTWO$ & $\BEmclfiSIXTWO$ \\
\hline
$\mclfiSEVENTWO$ & $\EmclfiSEVENTWO$ & $\NPmclfiSEVENTWO$ & $\NUPmclfiSEVENTWO$ & $\BEmclfiSEVENTWO$ \\
\hline
$\mclfiEIGHTTWO$ & $\EmclfiEIGHTTWO$ & $\NPmclfiEIGHTTWO$ & $\NUPmclfiEIGHTTWO$ & $\BEmclfiEIGHTTWO$ \\
\hline
$\mclfiNINETWO$ & $\EmclfiNINETWO$ & $\NPmclfiNINETWO$ & $\NUPmclfiNINETWO$ & $\BEmclfiNINETWO$ \\
\hline
$\mclfiONEZEROTWO$ & $\EmclfiONEZEROTWO$ & $\NPmclfiONEZEROTWO$ & $\NUPmclfiONEZEROTWO$ & $\BEmclfiONEZEROTWO$ \\
\hline
\end{baytabular}
\vfill
\columnbreak
\vspace*{\fill}
\noindent
\begin{baytabular}
$\pli$ & $\Epli$ & $\NPpli$ & $\NUPpli$ & $\BEpli$ \\
\hline\hline
$\mcpliONE$ & $\EmcpliONE$ & $\NPmcpliONE$ & $\NUPmcpliONE$ & $\BEmcpliONE$ \\
\hline
$\mcpliTWO$ & $\EmcpliTWO$ & $\NPmcpliTWO$ & $\NUPmcpliTWO$ & $\BEmcpliTWO$ \\
\hline
$\mcpliTHREE$ & $\EmcpliTHREE$ & $\NPmcpliTHREE$ & $\NUPmcpliTHREE$ & $\BEmcpliTHREE$ \\
\hline
$\mcpliFOUR$ & $\EmcpliFOUR$ & $\NPmcpliFOUR$ & $\NUPmcpliFOUR$ & $\BEmcpliFOUR$ \\
\hline
$\mcpliFIVE$ & $\EmcpliFIVE$ & $\NPmcpliFIVE$ & $\NUPmcpliFIVE$ & $\BEmcpliFIVE$ \\
\hline
$\mcpliSIX$ & $\EmcpliSIX$ & $\NPmcpliSIX$ & $\NUPmcpliSIX$ & $\BEmcpliSIX$ \\
\hline
$\mcpliSEVEN$ & $\EmcpliSEVEN$ & $\NPmcpliSEVEN$ & $\NUPmcpliSEVEN$ & $\BEmcpliSEVEN$ \\
\hline
$\mcpliEIGHT$ & $\EmcpliEIGHT$ & $\NPmcpliEIGHT$ & $\NUPmcpliEIGHT$ & $\BEmcpliEIGHT$ \\
\hline
$\mcpliNINE$ & $\EmcpliNINE$ & $\NPmcpliNINE$ & $\NUPmcpliNINE$ & $\BEmcpliNINE$ \\
\hline
$\mcpliONEZERO$ & $\EmcpliONEZERO$ & $\NPmcpliONEZERO$ & $\NUPmcpliONEZERO$ & $\BEmcpliONEZERO$ \\
\hline
\end{baytabular}
\end{multicols}
\end{center}
\caption{Bayesian analysis for the Large-Field (quadratic) Inflation and Power-Law Inflation models.}
\label{table:LFI2PLI}
\end{table}
}

\newcommand{\tableHINIAverage}{
\begin{table}[t]
\begin{center}
\begin{multicols}{2}
\noindent
\begin{baytabularAverage}
$\hi$ & & \multicolumn{3}{c|}{$\Ehi$}  \\
\hline\hline
$\mchiONE$ & $\SRmchiONE$ & $\EmchiONE$ & \multirow{3}{*}{$\EmchiA$} & \multirow{10}{*}{$\Emchi$} \\
\cline{1-3}
$\mchiTWO$ & $\SRmchiTWO$ & $\EmchiTWO$ &   &   \\
\cline{1-3}
$\mchiTHREE$ & $\SRmchiTHREE$ & $\EmchiTHREE$ &   &   \\
\cline{1-4}
$\mchiFOUR$ & $\SRmchiFOUR$ & $\EmchiFOUR$ & \multirow{3}{*}{$\EmchiB$} &  \\
\cline{1-3}
$\mchiFIVE$ & $\SRmchiFIVE$ & $\EmchiFIVE$ &   &   \\
\cline{1-3}
$\mchiSIX$ & $\SRmchiSIX$ & $\EmchiSIX$ &   &   \\
\cline{1-4}
$\mchiSEVEN$ & $\SRmchiSEVEN$ & $\EmchiSEVEN$ & \multirow{4}{*}{$\EmchiC$} &\\
\cline{1-3}
$\mchiEIGHT$ & $\SRmchiEIGHT$ & $\EmchiEIGHT$ &   &   \\
\cline{1-3}
$\mchiNINE$ & $\SRmchiNINE$ & $\EmchiNINE$ &   &   \\
\cline{1-3}
$\mchiONEZERO$ & $\SRmchiONEZERO$ & $\EmchiONEZERO$ &   &   \\
\hline
\end{baytabularAverage}
\vfill
\columnbreak
\vspace*{\fill}
\noindent
\begin{baytabularAverage}
$\nati$ & & \multicolumn{3}{c|}{$\Eni$}  \\
\hline\hline
$\mcniONE$ & $\SRmcniONE$ & $\EmcniONE$ & \multirow{3}{*}{$\EmcniA$} & \multirow{10}{*}{$\Emcni$} \\
\cline{1-3}
$\mcniTWO$ & $\SRmcniTWO$ & $\EmcniTWO$ &   &   \\
\cline{1-3}
$\mcniTHREE$ & $\SRmcniTHREE$ & $\EmcniTHREE$ &   &   \\
\cline{1-4}
$\mcniFOUR$ & $\SRmcniFOUR$ & $\EmcniFOUR$ & \multirow{3}{*}{$\EmcniB$} &  \\
\cline{1-3}
$\mcniFIVE$ & $\SRmcniFIVE$ & $\EmcniFIVE$ &   &   \\
\cline{1-3}
$\mcniSIX$ & $\SRmcniSIX$ & $\EmcniSIX$ &   &   \\
\cline{1-4}
$\mcniSEVEN$ & $\SRmcniSEVEN$ & $\EmcniSEVEN$ & \multirow{4}{*}{$\EmcniC$} &\\
\cline{1-3}
$\mcniEIGHT$ & $\SRmcniEIGHT$ & $\EmcniEIGHT$ &   &   \\
\cline{1-3}
$\mcniNINE$ & $\SRmcniNINE$ & $\EmcniNINE$ &   &   \\
\cline{1-3}
$\mcniONEZERO$ & $\SRmcniONEZERO$ & $\EmcniONEZERO$ &   &   \\
\hline
\end{baytabularAverage}
\end{multicols}
\end{center}
\caption{Prior volume fraction ($\%$) and averaged Bayesian evidence for the Higgs Inflation and Natural Inflation models.}
\label{table:HINI:Average}
\end{table}
}

\newcommand{\tableLFITWOPLIAverage}{
\begin{table}[t]
\begin{center}
\begin{multicols}{2}
\noindent
\begin{baytabularAverage}
$\lfiTWO$ & & \multicolumn{3}{c|}{$\ElfiTWO$}  \\
\hline\hline
$\mclfiONETWO$ & $\SRmclfiONETWO$ & $\EmclfiONETWO$ & \multirow{3}{*}{$\EmclfiTWOA$} & \multirow{10}{*}{$\EmclfiTWOG$} \\
\cline{1-3}
$\mclfiTWOTWO$ & $\SRmclfiTWOTWO$ & $\EmclfiTWOTWO$ &   &   \\
\cline{1-3}
$\mclfiTHREETWO$ & $\SRmclfiTHREETWO$ & $\EmclfiTHREETWO$ &   &   \\
\cline{1-4}
$\mclfiFOURTWO$ & $\SRmclfiFOURTWO$ & $\EmclfiFOURTWO$ & \multirow{3}{*}{$\EmclfiTWOB$} &  \\
\cline{1-3}
$\mclfiFIVETWO$ & $\SRmclfiFIVETWO$ & $\EmclfiFIVETWO$ &   &   \\
\cline{1-3}
$\mclfiSIXTWO$ & $\SRmclfiSIXTWO$ & $\EmclfiSIXTWO$ &   &   \\
\cline{1-4}
$\mclfiSEVENTWO$ & $\SRmclfiSEVENTWO$ & $\EmclfiSEVENTWO$ & \multirow{4}{*}{$\EmclfiTWOC$} &\\
\cline{1-3}
$\mclfiEIGHTTWO$ & $\SRmclfiEIGHTTWO$ & $\EmclfiEIGHTTWO$ &   &   \\
\cline{1-3}
$\mclfiNINETWO$ & $\SRmclfiNINETWO$ & $\EmclfiNINETWO$ &   &   \\
\cline{1-3}
$\mclfiONEZEROTWO$ & $\SRmclfiONEZEROTWO$ & $\EmclfiONEZEROTWO$ &   &   \\
\hline
\end{baytabularAverage}
\vfill
\columnbreak
\vspace*{\fill}
\noindent
\begin{baytabularAverage}
$\pli$ & & \multicolumn{3}{c|}{$\Epli$}  \\
\hline\hline
$\mcpliONE$ & $\SRmcpliONE$ & $\EmcpliONE$ & \multirow{3}{*}{$\EmcpliA$} & \multirow{10}{*}{$\Emcpli$} \\
\cline{1-3}
$\mcpliTWO$ & $\SRmcpliTWO$ & $\EmcpliTWO$ &   &   \\
\cline{1-3}
$\mcpliTHREE$ & $\SRmcpliTHREE$ & $\EmcpliTHREE$ &   &   \\
\cline{1-4}
$\mcpliFOUR$ & $\SRmcpliFOUR$ & $\EmcpliFOUR$ & \multirow{3}{*}{$\EmcpliB$} &  \\
\cline{1-3}
$\mcpliFIVE$ & $\SRmcpliFIVE$ & $\EmcpliFIVE$ &   &   \\
\cline{1-3}
$\mcpliSIX$ & $\SRmcpliSIX$ & $\EmcpliSIX$ &   &   \\
\cline{1-4}
$\mcpliSEVEN$ & $\SRmcpliSEVEN$ & $\EmcpliSEVEN$ & \multirow{4}{*}{$\EmcpliC$} &\\
\cline{1-3}
$\mcpliEIGHT$ & $\SRmcpliEIGHT$ & $\EmcpliEIGHT$ &   &   \\
\cline{1-3}
$\mcpliNINE$ & $\SRmcpliNINE$ & $\EmcpliNINE$ &   &   \\
\cline{1-3}
$\mcpliONEZERO$ & $\SRmcpliONEZERO$ & $\EmcpliONEZERO$ &   &   \\
\hline
\end{baytabularAverage}
\end{multicols}
\end{center}
\caption{Prior volume fraction ($\%$) and averaged Bayesian evidence for the Large-Field (quadratic) Inflation and Power-Law Inflation models.}
\label{table:LFI2PLI:Average}
\end{table}
}

\newcommand{\tableHINIAverageGaussian}{
\begin{table}[t]
\begin{center}
\begin{multicols}{2}
\noindent
\linebreak
\begin{baytabularAverage}
$\hi$ & & \multicolumn{3}{c|}{$\Ehi$}  \\
\hline\hline
$\mchiONE$ & $\SRmchiONEGaussian$ & $\EmchiONEGaussian$ & \multirow{3}{*}{$\EmchiAGaussian$} & \multirow{10}{*}{$\EmchiGaussian$} \\
\cline{1-3}
$\mchiTWO$ & $\SRmchiTWOGaussian$ & $\EmchiTWOGaussian$ &   &   \\
\cline{1-3}
$\mchiTHREE$ & $\SRmchiTHREEGaussian$ & $\EmchiTHREEGaussian$ &   &   \\
\cline{1-4}
$\mchiFOUR$ & $\SRmchiFOURGaussian$ & $\EmchiFOURGaussian$ & \multirow{3}{*}{$\EmchiBGaussian$} &  \\
\cline{1-3}
$\mchiFIVE$ & $\SRmchiFIVEGaussian$ & $\EmchiFIVEGaussian$ &   &   \\
\cline{1-3}
$\mchiSIX$ & $\SRmchiSIXGaussian$ & $\EmchiSIXGaussian$ &   &   \\
\cline{1-4}
$\mchiSEVEN$ & $\SRmchiSEVENGaussian$ & $\EmchiSEVENGaussian$ & \multirow{4}{*}{$\EmchiCGaussian$} &\\
\cline{1-3}
$\mchiEIGHT$ & $\SRmchiEIGHTGaussian$ & $\EmchiEIGHTGaussian$ &   &   \\
\cline{1-3}
$\mchiNINE$ & $\SRmchiNINEGaussian$ & $\EmchiNINEGaussian$ &   &   \\
\cline{1-3}
$\mchiONEZERO$ & $\SRmchiONEZEROGaussian$ & $\EmchiONEZEROGaussian$ &   &   \\
\hline
\end{baytabularAverage}
\vfill
\columnbreak
\vspace*{\fill}
\noindent
\begin{baytabularAverage}
$\nati$ & & \multicolumn{3}{c|}{$\Eni$}  \\
\hline\hline
$\mcniONE$ & $\SRmcniONEGaussian$ & $\EmcniONEGaussian$ & \multirow{3}{*}{$\EmcniAGaussian$} & \multirow{10}{*}{$\EmcniGaussian$} \\
\cline{1-3}
$\mcniTWO$ & $\SRmcniTWOGaussian$ & $\EmcniTWOGaussian$ &   &   \\
\cline{1-3}
$\mcniTHREE$ & $\SRmcniTHREEGaussian$ & $\EmcniTHREEGaussian$ &   &   \\
\cline{1-4}
$\mcniFOUR$ & $\SRmcniFOURGaussian$ & $\EmcniFOURGaussian$ & \multirow{3}{*}{$\EmcniBGaussian$} &  \\
\cline{1-3}
$\mcniFIVE$ & $\SRmcniFIVEGaussian$ & $\EmcniFIVEGaussian$ &   &   \\
\cline{1-3}
$\mcniSIX$ & $\SRmcniSIXGaussian$ & $\EmcniSIXGaussian$ &   &   \\
\cline{1-4}
$\mcniSEVEN$ & $\SRmcniSEVENGaussian$ & $\EmcniSEVENGaussian$ & \multirow{4}{*}{$\EmcniCGaussian$} &\\
\cline{1-3}
$\mcniEIGHT$ & $\SRmcniEIGHTGaussian$ & $\EmcniEIGHTGaussian$ &   &   \\
\cline{1-3}
$\mcniNINE$ & $\SRmcniNINEGaussian$ & $\EmcniNINEGaussian$ &   &   \\
\cline{1-3}
$\mcniONEZERO$ & $\SRmcniONEZEROGaussian$ & $\EmcniONEZEROGaussian$ &   &   \\
\hline
\end{baytabularAverage}
\end{multicols}
\end{center}
\caption{Prior volume fraction ($\%$) and averaged Bayesian evidence for the Higgs Inflation and Natural Inflation models, when a ``Gaussian'' prior is used for $\sigma_\uend$.}
\label{table:HINI:AverageGaussian}
\end{table}
}

\newcommand{\tableLFITWOPLIAverageGaussian}{
\begin{table}[t]
\begin{center}
\begin{multicols}{2}
\noindent
\linebreak
\begin{baytabularAverage}
$\lfiTWO$ & & \multicolumn{3}{c|}{$\ElfiTWO$}  \\
\hline\hline
$\mclfiONETWO$ & $\SRmclfiONETWOGaussian$ & $\EmclfiONETWOGaussian$ & \multirow{3}{*}{$\EmclfiTWOAGaussian$} & \multirow{10}{*}{$\EmclfiTWOGGaussian$} \\
\cline{1-3}
$\mclfiTWOTWO$ & $\SRmclfiTWOTWOGaussian$ & $\EmclfiTWOTWOGaussian$ &   &   \\
\cline{1-3}
$\mclfiTHREETWO$ & $\SRmclfiTHREETWOGaussian$ & $\EmclfiTHREETWOGaussian$ &   &   \\
\cline{1-4}
$\mclfiFOURTWO$ & $\SRmclfiFOURTWOGaussian$ & $\EmclfiFOURTWOGaussian$ & \multirow{3}{*}{$\EmclfiTWOBGaussian$} &  \\
\cline{1-3}
$\mclfiFIVETWO$ & $\SRmclfiFIVETWOGaussian$ & $\EmclfiFIVETWOGaussian$ &   &   \\
\cline{1-3}
$\mclfiSIXTWO$ & $\SRmclfiSIXTWOGaussian$ & $\EmclfiSIXTWOGaussian$ &   &   \\
\cline{1-4}
$\mclfiSEVENTWO$ & $\SRmclfiSEVENTWOGaussian$ & $\EmclfiSEVENTWOGaussian$ & \multirow{4}{*}{$\EmclfiTWOCGaussian$} &\\
\cline{1-3}
$\mclfiEIGHTTWO$ & $\SRmclfiEIGHTTWOGaussian$ & $\EmclfiEIGHTTWOGaussian$ &   &   \\
\cline{1-3}
$\mclfiNINETWO$ & $\SRmclfiNINETWOGaussian$ & $\EmclfiNINETWOGaussian$ &   &   \\
\cline{1-3}
$\mclfiONEZEROTWO$ & $\SRmclfiONEZEROTWOGaussian$ & $\EmclfiONEZEROTWOGaussian$ &   &   \\
\hline
\end{baytabularAverage}
\vfill
\columnbreak
\vspace*{\fill}
\noindent
\begin{baytabularAverage}
$\pli$ & & \multicolumn{3}{c|}{$\Epli$}  \\
\hline\hline
$\mcpliONE$ & $\SRmcpliONEGaussian$ & $\EmcpliONEGaussian$ & \multirow{3}{*}{$\EmcpliAGaussian$} & \multirow{10}{*}{$\EmcpliGaussian$} \\
\cline{1-3}
$\mcpliTWO$ & $\SRmcpliTWOGaussian$ & $\EmcpliTWOGaussian$ &   &   \\
\cline{1-3}
$\mcpliTHREE$ & $\SRmcpliTHREEGaussian$ & $\EmcpliTHREEGaussian$ &   &   \\
\cline{1-4}
$\mcpliFOUR$ & $\SRmcpliFOURGaussian$ & $\EmcpliFOURGaussian$ & \multirow{3}{*}{$\EmcpliBGaussian$} &  \\
\cline{1-3}
$\mcpliFIVE$ & $\SRmcpliFIVEGaussian$ & $\EmcpliFIVEGaussian$ &   &   \\
\cline{1-3}
$\mcpliSIX$ & $\SRmcpliSIXGaussian$ & $\EmcpliSIXGaussian$ &   &   \\
\cline{1-4}
$\mcpliSEVEN$ & $\SRmcpliSEVENGaussian$ & $\EmcpliSEVENGaussian$ & \multirow{4}{*}{$\EmcpliCGaussian$} &\\
\cline{1-3}
$\mcpliEIGHT$ & $\SRmcpliEIGHTGaussian$ & $\EmcpliEIGHTGaussian$ &   &   \\
\cline{1-3}
$\mcpliNINE$ & $\SRmcpliNINEGaussian$ & $\EmcpliNINEGaussian$ &   &   \\
\cline{1-3}
$\mcpliONEZERO$ & $\SRmcpliONEZEROGaussian$ & $\EmcpliONEZEROGaussian$ &   &   \\
\hline
\end{baytabularAverage}
\end{multicols}
\end{center}
\caption{Prior volume fraction ($\%$) and averaged Bayesian evidence for the Large-Field (quadratic) Inflation and Power-Law Inflation models, when a ``Gaussian'' prior is used for $\sigma_\uend$.}
\label{table:LFI2PLI:AverageGaussian}
\end{table}
}

\newcommand{\tableLFIAverageONE}{
\begin{multicols}{2}
\noindent
\linebreak
\begin{baytabularAverage}
$\lfi$ & & \multicolumn{3}{c|}{$\Elfi$}  \\
\hline\hline
$\mclfiONE$ & $\SRmclfiONE$ & $\EmclfiONE$ & \multirow{3}{*}{$\EmclfiA$} & \multirow{10}{*}{$\EmclfiG$} \\
\cline{1-3}
$\mclfiTWO$ & $\SRmclfiTWO$ & $\EmclfiTWO$ &   &   \\
\cline{1-3}
$\mclfiTHREE$ & $\SRmclfiTHREE$ & $\EmclfiTHREE$ &   &   \\
\cline{1-4}
$\mclfiFOUR$ & $\SRmclfiFOUR$ & $\EmclfiFOUR$ & \multirow{3}{*}{$\EmclfiB$} &  \\
\cline{1-3}
$\mclfiFIVE$ & $\SRmclfiFIVE$ & $\EmclfiFIVE$ &   &   \\
\cline{1-3}
$\mclfiSIX$ & $\SRmclfiSIX$ & $\EmclfiSIX$ &   &   \\
\cline{1-4}
$\mclfiSEVEN$ & $\SRmclfiSEVEN$ & $\EmclfiSEVEN$ & \multirow{4}{*}{$\EmclfiC$} &\\
\cline{1-3}
$\mclfiEIGHT$ & $\SRmclfiEIGHT$ & $\EmclfiEIGHT$ &   &   \\
\cline{1-3}
$\mclfiNINE$ & $\SRmclfiNINE$ & $\EmclfiNINE$ &   &   \\
\cline{1-3}
$\mclfiONEZERO$ & $\SRmclfiONEZERO$ & $\EmclfiONEZERO$ &   &   \\
\hline
\end{baytabularAverage}
\vfill
\columnbreak
\vspace*{\fill}
\noindent
\begin{baytabularAverage}
$\lfiTWOTHREE$ & & \multicolumn{3}{c|}{$\ElfiTWOTHREE$}  \\
\hline\hline
$\mclfiONETWOTHREE$ & $\SRmclfiONETWOTHREE$ & $\EmclfiONETWOTHREE$ & \multirow{3}{*}{$\EmclfiTWOTHREEA$} & \multirow{10}{*}{$\EmclfiTWOTHREEG$} \\
\cline{1-3}
$\mclfiTWOTWOTHREE$ & $\SRmclfiTWOTWOTHREE$ & $\EmclfiTWOTWOTHREE$ &   &   \\
\cline{1-3}
$\mclfiTHREETWOTHREE$ & $\SRmclfiTHREETWOTHREE$ & $\EmclfiTHREETWOTHREE$ &   &   \\
\cline{1-4}
$\mclfiFOURTWOTHREE$ & $\SRmclfiFOURTWOTHREE$ & $\EmclfiFOURTWOTHREE$ & \multirow{3}{*}{$\EmclfiTWOTHREEB$} &  \\
\cline{1-3}
$\mclfiFIVETWOTHREE$ & $\SRmclfiFIVETWOTHREE$ & $\EmclfiFIVETWOTHREE$ &   &   \\
\cline{1-3}
$\mclfiSIXTWOTHREE$ & $\SRmclfiSIXTWOTHREE$ & $\EmclfiSIXTWOTHREE$ &   &   \\
\cline{1-4}
$\mclfiSEVENTWOTHREE$ & $\SRmclfiSEVENTWOTHREE$ & $\EmclfiSEVENTWOTHREE$ & \multirow{4}{*}{$\EmclfiTWOTHREEC$} &\\
\cline{1-3}
$\mclfiEIGHTTWOTHREE$ & $\SRmclfiEIGHTTWOTHREE$ & $\EmclfiEIGHTTWOTHREE$ &   &   \\
\cline{1-3}
$\mclfiNINETWOTHREE$ & $\SRmclfiNINETWOTHREE$ & $\EmclfiNINETWOTHREE$ &   &   \\
\cline{1-3}
$\mclfiONEZEROTWOTHREE$ & $\SRmclfiONEZEROTWOTHREE$ & $\EmclfiONEZEROTWOTHREE$ &   &   \\
\hline
\end{baytabularAverage}
\end{multicols}
}

\newcommand{\tableLFIAverageTWO}{
\begin{multicols}{2}
\noindent
\linebreak
\begin{baytabularAverage}
$\lfiONE$ & & \multicolumn{3}{c|}{$\ElfiONE$}  \\
\hline\hline
$\mclfiONEONE$ & $\SRmclfiONEONE$ & $\EmclfiONEONE$ & \multirow{3}{*}{$\EmclfiONEA$} & \multirow{10}{*}{$\EmclfiONEG$} \\
\cline{1-3}
$\mclfiTWOONE$ & $\SRmclfiTWOONE$ & $\EmclfiTWOONE$ &   &   \\
\cline{1-3}
$\mclfiTHREEONE$ & $\SRmclfiTHREEONE$ & $\EmclfiTHREEONE$ &   &   \\
\cline{1-4}
$\mclfiFOURONE$ & $\SRmclfiFOURONE$ & $\EmclfiFOURONE$ & \multirow{3}{*}{$\EmclfiONEB$} &  \\
\cline{1-3}
$\mclfiFIVEONE$ & $\SRmclfiFIVEONE$ & $\EmclfiFIVEONE$ &   &   \\
\cline{1-3}
$\mclfiSIXONE$ & $\SRmclfiSIXONE$ & $\EmclfiSIXONE$ &   &   \\
\cline{1-4}
$\mclfiSEVENONE$ & $\SRmclfiSEVENONE$ & $\EmclfiSEVENONE$ & \multirow{4}{*}{$\EmclfiONEC$} &\\
\cline{1-3}
$\mclfiEIGHTONE$ & $\SRmclfiEIGHTONE$ & $\EmclfiEIGHTONE$ &   &   \\
\cline{1-3}
$\mclfiNINEONE$ & $\SRmclfiNINEONE$ & $\EmclfiNINEONE$ &   &   \\
\cline{1-3}
$\mclfiONEZEROONE$ & $\SRmclfiONEZEROONE$ & $\EmclfiONEZEROONE$ &   &   \\
\hline
\end{baytabularAverage}
\vfill
\columnbreak
\vspace*{\fill}
\noindent
\begin{baytabularAverage}
$\lfiTWO$ & & \multicolumn{3}{c|}{$\ElfiTWO$}  \\
\hline\hline
$\mclfiONETWO$ & $\SRmclfiONETWO$ & $\EmclfiONETWO$ & \multirow{3}{*}{$\EmclfiTWOA$} & \multirow{10}{*}{$\EmclfiTWOG$} \\
\cline{1-3}
$\mclfiTWOTWO$ & $\SRmclfiTWOTWO$ & $\EmclfiTWOTWO$ &   &   \\
\cline{1-3}
$\mclfiTHREETWO$ & $\SRmclfiTHREETWO$ & $\EmclfiTHREETWO$ &   &   \\
\cline{1-4}
$\mclfiFOURTWO$ & $\SRmclfiFOURTWO$ & $\EmclfiFOURTWO$ & \multirow{3}{*}{$\EmclfiTWOB$} &  \\
\cline{1-3}
$\mclfiFIVETWO$ & $\SRmclfiFIVETWO$ & $\EmclfiFIVETWO$ &   &   \\
\cline{1-3}
$\mclfiSIXTWO$ & $\SRmclfiSIXTWO$ & $\EmclfiSIXTWO$ &   &   \\
\cline{1-4}
$\mclfiSEVENTWO$ & $\SRmclfiSEVENTWO$ & $\EmclfiSEVENTWO$ & \multirow{4}{*}{$\EmclfiTWOC$} &\\
\cline{1-3}
$\mclfiEIGHTTWO$ & $\SRmclfiEIGHTTWO$ & $\EmclfiEIGHTTWO$ &   &   \\
\cline{1-3}
$\mclfiNINETWO$ & $\SRmclfiNINETWO$ & $\EmclfiNINETWO$ &   &   \\
\cline{1-3}
$\mclfiONEZEROTWO$ & $\SRmclfiONEZEROTWO$ & $\EmclfiONEZEROTWO$ &   &   \\
\hline
\end{baytabularAverage}
\end{multicols}
}

\newcommand{\tableLFIAverageTHREE}{
\begin{multicols}{2}
\noindent
\linebreak
\begin{baytabularAverage}
$\lfiTHREE$ & & \multicolumn{3}{c|}{$\ElfiTHREE$}  \\
\hline\hline
$\mclfiONETHREE$ & $\SRmclfiONETHREE$ & $\EmclfiONETHREE$ & \multirow{3}{*}{$\EmclfiTHREEA$} & \multirow{10}{*}{$\EmclfiTHREEG$} \\
\cline{1-3}
$\mclfiTWOTHREE$ & $\SRmclfiTWOTHREE$ & $\EmclfiTWOTHREE$ &   &   \\
\cline{1-3}
$\mclfiTHREETHREE$ & $\SRmclfiTHREETHREE$ & $\EmclfiTHREETHREE$ &   &   \\
\cline{1-4}
$\mclfiFOURTHREE$ & $\SRmclfiFOURTHREE$ & $\EmclfiFOURTHREE$ & \multirow{3}{*}{$\EmclfiTHREEB$} &  \\
\cline{1-3}
$\mclfiFIVETHREE$ & $\SRmclfiFIVETHREE$ & $\EmclfiFIVETHREE$ &   &   \\
\cline{1-3}
$\mclfiSIXTHREE$ & $\SRmclfiSIXTHREE$ & $\EmclfiSIXTHREE$ &   &   \\
\cline{1-4}
$\mclfiSEVENTHREE$ & $\SRmclfiSEVENTHREE$ & $\EmclfiSEVENTHREE$ & \multirow{4}{*}{$\EmclfiTHREEC$} &\\
\cline{1-3}
$\mclfiEIGHTTHREE$ & $\SRmclfiEIGHTTHREE$ & $\EmclfiEIGHTTHREE$ &   &   \\
\cline{1-3}
$\mclfiNINETHREE$ & $\SRmclfiNINETHREE$ & $\EmclfiNINETHREE$ &   &   \\
\cline{1-3}
$\mclfiONEZEROTHREE$ & $\SRmclfiONEZEROTHREE$ & $\EmclfiONEZEROTHREE$ &   &   \\
\hline
\end{baytabularAverage}
\vfill
\columnbreak
\vspace*{\fill}
\noindent
\begin{baytabularAverage}
$\lfiFOUR$ & & \multicolumn{3}{c|}{$\ElfiFOUR$}  \\
\hline\hline
$\mclfiONEFOUR$ & $\SRmclfiONEFOUR$ & $\EmclfiONEFOUR$ & \multirow{3}{*}{$\EmclfiFOURA$} & \multirow{10}{*}{$\EmclfiFOURG$} \\
\cline{1-3}
$\mclfiTWOFOUR$ & $\SRmclfiTWOFOUR$ & $\EmclfiTWOFOUR$ &   &   \\
\cline{1-3}
$\mclfiTHREEFOUR$ & $\SRmclfiTHREEFOUR$ & $\EmclfiTHREEFOUR$ &   &   \\
\cline{1-4}
$\mclfiFOURFOUR$ & $\SRmclfiFOURFOUR$ & $\EmclfiFOURFOUR$ & \multirow{3}{*}{$\EmclfiFOURB$} &  \\
\cline{1-3}
$\mclfiFIVEFOUR$ & $\SRmclfiFIVEFOUR$ & $\EmclfiFIVEFOUR$ &   &   \\
\cline{1-3}
$\mclfiSIXFOUR$ & $\SRmclfiSIXFOUR$ & $\EmclfiSIXFOUR$ &   &   \\
\cline{1-4}
$\mclfiSEVENFOUR$ & $\SRmclfiSEVENFOUR$ & $\EmclfiSEVENFOUR$ & \multirow{4}{*}{$\EmclfiFOURC$} &\\
\cline{1-3}
$\mclfiEIGHTFOUR$ & $\SRmclfiEIGHTFOUR$ & $\EmclfiEIGHTFOUR$ &   &   \\
\cline{1-3}
$\mclfiNINEFOUR$ & $\SRmclfiNINEFOUR$ & $\EmclfiNINEFOUR$ &   &   \\
\cline{1-3}
$\mclfiONEZEROFOUR$ & $\SRmclfiONEZEROFOUR$ & $\EmclfiONEZEROFOUR$ &   &   \\
\hline
\end{baytabularAverage}
\end{multicols}
}

\newcommand{\tableLFIAverageONEGaussian}{
\begin{multicols}{2}
\noindent
\linebreak
\begin{baytabularAverage}
$\lfi$ & & \multicolumn{3}{c|}{$\Elfi$}  \\
\hline\hline
$\mclfiONE$ & $\SRmclfiONEGaussian$ & $\EmclfiONEGaussian$ & \multirow{3}{*}{$\EmclfiAGaussian$} & \multirow{10}{*}{$\EmclfiGGaussian$} \\
\cline{1-3}
$\mclfiTWO$ & $\SRmclfiTWOGaussian$ & $\EmclfiTWOGaussian$ &   &   \\
\cline{1-3}
$\mclfiTHREE$ & $\SRmclfiTHREEGaussian$ & $\EmclfiTHREEGaussian$ &   &   \\
\cline{1-4}
$\mclfiFOUR$ & $\SRmclfiFOURGaussian$ & $\EmclfiFOURGaussian$ & \multirow{3}{*}{$\EmclfiBGaussian$} &  \\
\cline{1-3}
$\mclfiFIVE$ & $\SRmclfiFIVEGaussian$ & $\EmclfiFIVEGaussian$ &   &   \\
\cline{1-3}
$\mclfiSIX$ & $\SRmclfiSIXGaussian$ & $\EmclfiSIXGaussian$ &   &   \\
\cline{1-4}
$\mclfiSEVEN$ & $\SRmclfiSEVENGaussian$ & $\EmclfiSEVENGaussian$ & \multirow{4}{*}{$\EmclfiCGaussian$} &\\
\cline{1-3}
$\mclfiEIGHT$ & $\SRmclfiEIGHTGaussian$ & $\EmclfiEIGHTGaussian$ &   &   \\
\cline{1-3}
$\mclfiNINE$ & $\SRmclfiNINEGaussian$ & $\EmclfiNINEGaussian$ &   &   \\
\cline{1-3}
$\mclfiONEZERO$ & $\SRmclfiONEZEROGaussian$ & $\EmclfiONEZEROGaussian$ &   &   \\
\hline
\end{baytabularAverage}
\vfill
\columnbreak
\vspace*{\fill}
\noindent
\begin{baytabularAverage}
$\lfiTWOTHREE$ & & \multicolumn{3}{c|}{$\ElfiTWOTHREE$}  \\
\hline\hline
$\mclfiONETWOTHREE$ & $\SRmclfiONETWOTHREEGaussian$ & $\EmclfiONETWOTHREEGaussian$ & \multirow{3}{*}{$\EmclfiTWOTHREEAGaussian$} & \multirow{10}{*}{$\EmclfiTWOTHREEGGaussian$} \\
\cline{1-3}
$\mclfiTWOTWOTHREE$ & $\SRmclfiTWOTWOTHREEGaussian$ & $\EmclfiTWOTWOTHREEGaussian$ &   &   \\
\cline{1-3}
$\mclfiTHREETWOTHREE$ & $\SRmclfiTHREETWOTHREEGaussian$ & $\EmclfiTHREETWOTHREEGaussian$ &   &   \\
\cline{1-4}
$\mclfiFOURTWOTHREE$ & $\SRmclfiFOURTWOTHREEGaussian$ & $\EmclfiFOURTWOTHREEGaussian$ & \multirow{3}{*}{$\EmclfiTWOTHREEBGaussian$} &  \\
\cline{1-3}
$\mclfiFIVETWOTHREE$ & $\SRmclfiFIVETWOTHREEGaussian$ & $\EmclfiFIVETWOTHREEGaussian$ &   &   \\
\cline{1-3}
$\mclfiSIXTWOTHREE$ & $\SRmclfiSIXTWOTHREEGaussian$ & $\EmclfiSIXTWOTHREEGaussian$ &   &   \\
\cline{1-4}
$\mclfiSEVENTWOTHREE$ & $\SRmclfiSEVENTWOTHREEGaussian$ & $\EmclfiSEVENTWOTHREEGaussian$ & \multirow{4}{*}{$\EmclfiTWOTHREECGaussian$} &\\
\cline{1-3}
$\mclfiEIGHTTWOTHREE$ & $\SRmclfiEIGHTTWOTHREEGaussian$ & $\EmclfiEIGHTTWOTHREEGaussian$ &   &   \\
\cline{1-3}
$\mclfiNINETWOTHREE$ & $\SRmclfiNINETWOTHREEGaussian$ & $\EmclfiNINETWOTHREEGaussian$ &   &   \\
\cline{1-3}
$\mclfiONEZEROTWOTHREE$ & $\SRmclfiONEZEROTWOTHREEGaussian$ & $\EmclfiONEZEROTWOTHREEGaussian$ &   &   \\
\hline
\end{baytabularAverage}
\end{multicols}
}

\newcommand{\tableLFIAverageTWOGaussian}{
\begin{multicols}{2}
\noindent
\linebreak
\begin{baytabularAverage}
$\lfiONE$ & & \multicolumn{3}{c|}{$\ElfiONE$}  \\
\hline\hline
$\mclfiONEONE$ & $\SRmclfiONEONEGaussian$ & $\EmclfiONEONEGaussian$ & \multirow{3}{*}{$\EmclfiONEAGaussian$} & \multirow{10}{*}{$\EmclfiONEGGaussian$} \\
\cline{1-3}
$\mclfiTWOONE$ & $\SRmclfiTWOONEGaussian$ & $\EmclfiTWOONEGaussian$ &   &   \\
\cline{1-3}
$\mclfiTHREEONE$ & $\SRmclfiTHREEONEGaussian$ & $\EmclfiTHREEONEGaussian$ &   &   \\
\cline{1-4}
$\mclfiFOURONE$ & $\SRmclfiFOURONEGaussian$ & $\EmclfiFOURONEGaussian$ & \multirow{3}{*}{$\EmclfiONEBGaussian$} &  \\
\cline{1-3}
$\mclfiFIVEONE$ & $\SRmclfiFIVEONEGaussian$ & $\EmclfiFIVEONEGaussian$ &   &   \\
\cline{1-3}
$\mclfiSIXONE$ & $\SRmclfiSIXONEGaussian$ & $\EmclfiSIXONEGaussian$ &   &   \\
\cline{1-4}
$\mclfiSEVENONE$ & $\SRmclfiSEVENONEGaussian$ & $\EmclfiSEVENONEGaussian$ & \multirow{4}{*}{$\EmclfiONECGaussian$} &\\
\cline{1-3}
$\mclfiEIGHTONE$ & $\SRmclfiEIGHTONEGaussian$ & $\EmclfiEIGHTONEGaussian$ &   &   \\
\cline{1-3}
$\mclfiNINEONE$ & $\SRmclfiNINEONEGaussian$ & $\EmclfiNINEONEGaussian$ &   &   \\
\cline{1-3}
$\mclfiONEZEROONE$ & $\SRmclfiONEZEROONEGaussian$ & $\EmclfiONEZEROONEGaussian$ &   &   \\
\hline
\end{baytabularAverage}
\vfill
\columnbreak
\vspace*{\fill}
\noindent
\begin{baytabularAverage}
$\lfiTWO$ & & \multicolumn{3}{c|}{$\ElfiTWO$}  \\
\hline\hline
$\mclfiONETWO$ & $\SRmclfiONETWOGaussian$ & $\EmclfiONETWOGaussian$ & \multirow{3}{*}{$\EmclfiTWOAGaussian$} & \multirow{10}{*}{$\EmclfiTWOGGaussian$} \\
\cline{1-3}
$\mclfiTWOTWO$ & $\SRmclfiTWOTWOGaussian$ & $\EmclfiTWOTWOGaussian$ &   &   \\
\cline{1-3}
$\mclfiTHREETWO$ & $\SRmclfiTHREETWOGaussian$ & $\EmclfiTHREETWOGaussian$ &   &   \\
\cline{1-4}
$\mclfiFOURTWO$ & $\SRmclfiFOURTWOGaussian$ & $\EmclfiFOURTWOGaussian$ & \multirow{3}{*}{$\EmclfiTWOBGaussian$} &  \\
\cline{1-3}
$\mclfiFIVETWO$ & $\SRmclfiFIVETWOGaussian$ & $\EmclfiFIVETWOGaussian$ &   &   \\
\cline{1-3}
$\mclfiSIXTWO$ & $\SRmclfiSIXTWOGaussian$ & $\EmclfiSIXTWOGaussian$ &   &   \\
\cline{1-4}
$\mclfiSEVENTWO$ & $\SRmclfiSEVENTWOGaussian$ & $\EmclfiSEVENTWOGaussian$ & \multirow{4}{*}{$\EmclfiTWOCGaussian$} &\\
\cline{1-3}
$\mclfiEIGHTTWO$ & $\SRmclfiEIGHTTWOGaussian$ & $\EmclfiEIGHTTWOGaussian$ &   &   \\
\cline{1-3}
$\mclfiNINETWO$ & $\SRmclfiNINETWOGaussian$ & $\EmclfiNINETWOGaussian$ &   &   \\
\cline{1-3}
$\mclfiONEZEROTWO$ & $\SRmclfiONEZEROTWOGaussian$ & $\EmclfiONEZEROTWOGaussian$ &   &   \\
\hline
\end{baytabularAverage}
\end{multicols}
}

\newcommand{\tableLFIAverageTHREEGaussian}{
\begin{multicols}{2}
\noindent
\linebreak
\begin{baytabularAverage}
$\lfiTHREE$ & & \multicolumn{3}{c|}{$\ElfiTHREE$}  \\
\hline\hline
$\mclfiONETHREE$ & $\SRmclfiONETHREEGaussian$ & $\EmclfiONETHREEGaussian$ & \multirow{3}{*}{$\EmclfiTHREEAGaussian$} & \multirow{10}{*}{$\EmclfiTHREEGGaussian$} \\
\cline{1-3}
$\mclfiTWOTHREE$ & $\SRmclfiTWOTHREEGaussian$ & $\EmclfiTWOTHREEGaussian$ &   &   \\
\cline{1-3}
$\mclfiTHREETHREE$ & $\SRmclfiTHREETHREEGaussian$ & $\EmclfiTHREETHREEGaussian$ &   &   \\
\cline{1-4}
$\mclfiFOURTHREE$ & $\SRmclfiFOURTHREEGaussian$ & $\EmclfiFOURTHREEGaussian$ & \multirow{3}{*}{$\EmclfiTHREEB$} &  \\
\cline{1-3}
$\mclfiFIVETHREE$ & $\SRmclfiFIVETHREEGaussian$ & $\EmclfiFIVETHREEGaussian$ &   &   \\
\cline{1-3}
$\mclfiSIXTHREE$ & $\SRmclfiSIXTHREEGaussian$ & $\EmclfiSIXTHREEGaussian$ &   &   \\
\cline{1-4}
$\mclfiSEVENTHREE$ & $\SRmclfiSEVENTHREEGaussian$ & $\EmclfiSEVENTHREEGaussian$ & \multirow{4}{*}{$\EmclfiTHREECGaussian$} &\\
\cline{1-3}
$\mclfiEIGHTTHREE$ & $\SRmclfiEIGHTTHREEGaussian$ & $\EmclfiEIGHTTHREEGaussian$ &   &   \\
\cline{1-3}
$\mclfiNINETHREE$ & $\SRmclfiNINETHREEGaussian$ & $\EmclfiNINETHREEGaussian$ &   &   \\
\cline{1-3}
$\mclfiONEZEROTHREE$ & $\SRmclfiONEZEROTHREEGaussian$ & $\EmclfiONEZEROTHREEGaussian$ &   &   \\
\hline
\end{baytabularAverage}
\vfill
\columnbreak
\vspace*{\fill}
\noindent
\begin{baytabularAverage}
$\lfiFOUR$ & & \multicolumn{3}{c|}{$\ElfiFOUR$}  \\
\hline\hline
$\mclfiONEFOUR$ & $\SRmclfiONEFOURGaussian$ & $\EmclfiONEFOURGaussian$ & \multirow{3}{*}{$\EmclfiFOURAGaussian$} & \multirow{10}{*}{$\EmclfiFOURGGaussian$} \\
\cline{1-3}
$\mclfiTWOFOUR$ & $\SRmclfiTWOFOURGaussian$ & $\EmclfiTWOFOURGaussian$ &   &   \\
\cline{1-3}
$\mclfiTHREEFOUR$ & $\SRmclfiTHREEFOURGaussian$ & $\EmclfiTHREEFOURGaussian$ &   &   \\
\cline{1-4}
$\mclfiFOURFOUR$ & $\SRmclfiFOURFOURGaussian$ & $\EmclfiFOURFOURGaussian$ & \multirow{3}{*}{$\EmclfiFOURBGaussian$} &  \\
\cline{1-3}
$\mclfiFIVEFOUR$ & $\SRmclfiFIVEFOURGaussian$ & $\EmclfiFIVEFOURGaussian$ &   &   \\
\cline{1-3}
$\mclfiSIXFOUR$ & $\SRmclfiSIXFOURGaussian$ & $\EmclfiSIXFOURGaussian$ &   &   \\
\cline{1-4}
$\mclfiSEVENFOUR$ & $\SRmclfiSEVENFOURGaussian$ & $\EmclfiSEVENFOURGaussian$ & \multirow{4}{*}{$\EmclfiFOURCGaussian$} &\\
\cline{1-3}
$\mclfiEIGHTFOUR$ & $\SRmclfiEIGHTFOURGaussian$ & $\EmclfiEIGHTFOURGaussian$ &   &   \\
\cline{1-3}
$\mclfiNINEFOUR$ & $\SRmclfiNINEFOURGaussian$ & $\EmclfiNINEFOURGaussian$ &   &   \\
\cline{1-3}
$\mclfiONEZEROFOUR$ & $\SRmclfiONEZEROFOURGaussian$ & $\EmclfiONEZEROFOURGaussian$ &   &   \\
\hline
\end{baytabularAverage}
\end{multicols}
}